# Revisiting the Edge of Chaos: Evolving Cellular Automata to Perform Computations

Melanie Mitchell[1], Peter T. Hraber[1], and James P. Crutchfield[2]

## Abstract

We present results from an experiment similar to one performed by Packard [23], in which a genetic algorithm is used to evolve cellular automata (CA) to perform a particular computational task. Packard examined the frequency of evolved CA rules as a function of Langton's $\lambda$ parameter [16], and interpreted the results of his experiment as giving evidence for the following two hypotheses: (1) CA rules able to perform complex computations are most likely to be found near "critical" $\lambda$ values, which have been claimed to correlate with a phase transition between ordered and chaotic behavioral regimes for CA; (2) When CA rules are evolved to perform a complex computation, evolution will tend to select rules with $\lambda$ values close to the critical values. Our experiment produced very different results, and we suggest that the interpretation of the original results is not correct. We also review and discuss issues related to $\lambda$, dynamical-behavior classes, and computation in CA.

The main constructive results of our study are identifying the emergence and competition of computational strategies and analyzing the central role of symmetries in an evolutionary system. In particular, we demonstrate how symmetry breaking can impede the evolution toward higher computational capability.

---

[1]Santa Fe Institute, 1660 Old Pecos Trail, Suite A, Santa Fe, New Mexico, U.S.A. 87501. Email: mm@santafe.edu, pth@santafe.edu

[2]Physics Department, University of California, Berkeley, CA, U.S.A. 94720. Email: chaos@gojira.berkeley.edu

## 1. Introduction

The notion of "computation at the edge of chaos" has gained considerable attention in the study of complex systems and artificial life (e.g., [3, 4, 14, 16, 23, 30]). This notion is related to the broad question, What is the relation between a computational system's ability for complex information processing and other measures of the system's behavior? In particular, does the ability for nontrivial computation require a system's dynamical behavior to be "near a transition to chaos"? There has also been considerable attention given to the notion of "the edge of chaos" in the context of evolution. In particular, it has been hypothesized that when biological systems must perform complex computation in order to survive, the process of evolution under natural selection tends to select such systems near a phase transition from ordered to chaotic behavior [13, 14, 23].

This paper describes a re-examination of one study that addressed these questions in the context of cellular automata [23]. The results of the original study were interpreted as evidence that an evolutionary process in which cellular-automata rules are selected to perform a nontrivial computation preferentially selected rules near the transition to chaos. We show that this conclusion is neither supported by our experimental results nor consistent with basic mathematical properties of the computation being evolved. In the process of this demonstration, we review and clarify notions relating to terms such as "computation", "dynamical behavior", and "edge of chaos" in the context of cellular automata.

## 2. Cellular Automata and Dynamics

Cellular automata (CA) are discrete spatially-extended dynamical systems that have been studied extensively as models of physical processes and as computational devices [6, 10, 25, 29, 31]. In its simplest form, a CA consists of a spatial lattice of *cells*, each of which, at time $t$, can be in one of $K$ states. We denote the lattice size or number of cells as $L$. A CA has a single fixed rule used to update each cell; the rule maps from the states in a neighborhood of cells—e.g., the states of a cell and its nearest neighbors—to a single state, which is the update value for the cell in question. The lattice starts out with some initial configuration of local states and, at each time step, the states of all cells in the lattice are synchronously updated. In the following we will use the term "state" to refer to the value of a single cell—e.g., 0 or 1—and "configuration" to mean the pattern of states over the entire lattice.

The CA we will discuss in this paper are all one-dimensional with two possible states per cell (0 and 1). In a one-dimensional CA, the neighborhood of a cell includes the cell itself and some number of neighbors on either side of the cell. The number of neighbors on either side of the center cell is referred to as the CA's *radius* $r$. All of the simulations will be of CA with spatially periodic boundary conditions (i.e., the one-dimensional lattice is viewed as a circle, with the right neighbor of the rightmost cell being the leftmost cell, and vice versa).

The equations of motion for a CA are often expressed in the form of a *rule table*. This is a look-up table listing each of the neighborhood patterns and the state to which the central cell in that neighborhood is mapped. For example, Figure 1 displays one possible rule table for an "elementary" one-dimensional two-state CA with radius $r = 1$. The left-hand column gives the 8 possible neighborhood configurations, and the states in the right-hand column

```
000  0
001  0
010  0
011  1
100  0
101  1
110  1
111  1
```

Figure 1: Rule table for the "elementary"—binary-state, nearest-neighbor—CA with rule number 232. The $8 = 2^3$ neighborhood patterns are given in the left-hand column. For each of these the right-hand column gives the "output bit", the center cell's value at the next time step.

are referred to as the "output bits" of the rule table. To run the CA, this look-up table is applied to each neighborhood in the current lattice configuration, respecting the choice of boundary conditions, to produce the configuration at the next time step.

A common method for examining the behavior of a two-state one-dimensional CA is to display its space-time diagram, a two-dimensional picture that vertically strings together the one-dimensional CA lattice configurations at each successive time step, with white squares corresponding to cells in state 0, and black squares corresponding to cells in state 1. Two such space-time diagrams are displayed in Figure 2. These show the actions of the Gacs-Kurdyumov-Levin (GKL) binary-state CA on two random initial configurations of different densities of 1's [5, 7]. In both cases, over time the CA relaxes to a fixed pattern—in one case, all 0's, and in the other case, all 1's. These patterns are, in fact, fixed points of the GKL CA. That is, once reached, further applications of the CA do not change the pattern. The GKL CA will be discussed further below.

CA are of interest as models of physical processes because, like many physical systems, they consist of a large number of simple components (cells) which are modified only by local interactions, but which acting together can produce global complex behavior. Like the class of dissipative dynamical systems, even the class of elementary one-dimensional CA exhibit the full spectrum of dynamical behavior: from fixed points, as seen in Figure 2, to limit cycles (periodic behavior) to unpredictable ("chaotic") behavior. Wolfram considered a coarse classification of CA behavior in terms of these categories. He proposed the following four classes with the intention of capturing all possible CA behavior [30]:

*Class 1:* All initial configurations relax after a transient period to the same fixed configuration (e.g., all 1's).

*Class 2:* All initial configurations relax after a transient period to some fixed point or some temporally periodic cycle of configurations, but which one depends on the initial configuration.

*Class 3:* Some initial configurations relax after a transient period to chaotic behavior.

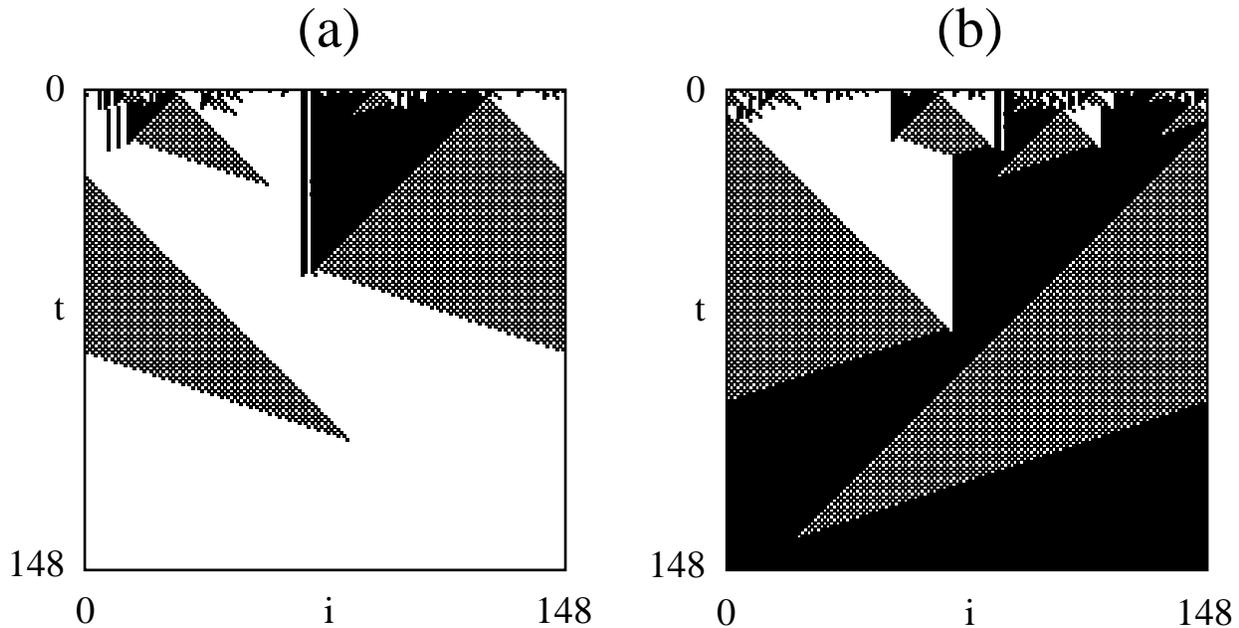

Figure 2: Two space-time diagrams for the binary-state Gacs-Kurdyumov-Levin CA. $L = 149$ sites are shown evolving, with time increasing down the page, from two different initial configurations over 149 time steps. In (a) the initial configuration has a density of 1's of approximately 0.48; in (b) a density of approximately 0.52. Notice that by the last time step the CA has converged to a fixed pattern of (a) all 0's and (b) all 1's. In this way the CA has classified the initial configurations according to their density.

(The term "chaotic" here and in the rest of this paper refers to apparently unpredictable space-time behavior.)

*Class 4:* Some initial configurations result in complex localized structures, sometimes long-lived.

Wolfram does not state the requirements for membership in Class 4 any more precisely than is given above. Thus, unlike the categories derived from dynamical systems theory, Class 4 is not rigorously defined.

It should be pointed out that on finite lattices, there is only a finite number ($2^L$) of possible configurations, so all rules ultimately lead to periodic behavior. Class 2 refers not to this type of periodic behavior but rather to cycles with periods much shorter than $2^L$.

## 3. Cellular Automata and Computation

CA are also of interest as computational devices, both as theoretical tools and as practical highly efficient parallel machines [25, 26, 29, 31].

"Computation" in the context of CA has several possible meanings. The most common meaning is that the CA does some "useful" computational task. Here, the rule is interpreted as the "program", the initial configuration is interpreted as the "input", and the CA runs for some specified number of time steps or until it reaches some "goal" pattern—possibly a fixed point pattern. The final pattern is interpreted as the "output". An example of this is using CA to perform image-processing tasks [26].

A second meaning of computation in CA is for a CA, given certain special initial configurations, to be capable of universal computation. That is, the CA can, given the right initial configuration, simulate a programmable computer, complete with logical gates, timing devices, and so on. Conway's Game of Life [1] is such a CA; one construction for universal computation in the Game of Life is given in [1]. Similar constructions have been made for one-dimensional CA [20]. Wolfram speculated that all Class 4 rules have the capacity for universal computation [30]. However, given the informality of the definition of Class 4, not to mention the difficulty of proving that a given rule is or is not capable of universal computation, this hypothesis is impossible to verify.

A third meaning of computation in CA involves interpreting the behavior of a given CA on an ensemble of initial configurations as a kind of "intrinsic" computation. Here computation is not interpreted as the performance of a "useful" transformation of the input to produce the output. Rather, it is measured in terms of generic, structural computational elements such as memory, information production, information transfer, logical operations, and so on. It is important to emphasize that the measurement of such intrinsic computational elements does not rely on a semantics of utility as do the preceding computation types. That is, these elements can be detected and quantified without reference to any specific "useful" computation performed by the CA—such as enhancing edges in an image or computing the digits of $\pi$. This notion of intrinsic computation is central to the work of Crutchfield, Hanson, and Young [3, 11].

Generally, CA have both the capacity for all kinds of dynamical behaviors and the capacity for all kinds of computational behaviors. For these reasons, in addition to the computational ease of simulating them, CA have been considered a good class of models to use in studying how dynamical behavior and computational ability are related. Similar questions have also been addressed in the context of other dynamical systems, including continuous-state dynamical systems such as iterated maps and differential equations [3, 4], Boolean networks [13], and recurrent neural networks [24]. Here we will confine our discussion to CA.

With this background, we can now rephrase the broad questions presented in Section 1 in the context of CA:

- What properties must a CA have for nontrivial computation?

- In particular, does a capacity for nontrivial computation, in any of the three senses described above, require a CA to be near a transition from ordered to chaotic behavior?

- When CA rules are evolved to perform a nontrivial computation, will evolution tend to select rules near such a transition to chaos?

4. **Structure of CA Rule Space**

Over the last decade there have been a number of studies addressing the first question above. Here we focus on Langton's empirical investigations of the second question in terms of the structure of the space of CA rules [16]. The relationship of the first two questions to the third—evolving CA—will be described subsequently.

One of the major difficulties in understanding the structure of the space of CA rules and its relation to computational capability is its discrete nature. In contrast to the well-developed theory of bifurcations for continuous-state dynamical systems[9], there appears to be little or no geometry in CA space and there is no notion of smoothly changing one CA to get another "nearby in behavior". In an attempt to emulate this, however, Langton defined a parameter $\lambda$ that varies incrementally as single output bits are turned on or off in a given rule table. For a given CA rule table, $\lambda$ is computed as follows. For a $K$-state CA, one state $q$ is chosen arbitrarily to be "quiescent".[3] The $\lambda$ of a given CA rule is then the fraction of non-quiescent output states in the rule table. For a binary-state CA, if 0 is chosen to be the quiescent state, then $\lambda$ is simply the fraction of output 1 bits in the rule table. Typically there are many CA rules with a given $\lambda$ value. For a binary CA, the number is strongly peaked at $\lambda = 1/2$, due to the combinatorial dependence on the radius $r$ and the number of states $K$. It is also symmetric about $\lambda = 1/2$, due to the symmetry of exchanging 0's and 1's. Generally, as $\lambda$ is increased from 0 to $[1 - 1/K]$, the CA move from having the most homogeneous rule tables to having the most heterogeneous.

Langton performed a range of Monte Carlo samples of two-dimensional CA in an attempt to characterize their average behavior as a function of $\lambda$ [16]. The notion of "average behavior" was intended to capture the most likely behavior observed with a randomly chosen initial configuration for CA randomly selected in a fixed-$\lambda$ subspace. The observation was that as $\lambda$ is incremented from 0 to $[1 - 1/K]$ the average behavior of rules passes through the following regimes:

$$\text{fixed point} \Rightarrow \text{periodic} \Rightarrow \text{``complex''} \Rightarrow \text{chaotic}.$$

That is, according to Figure 16 in [16], for example, the average behavior at low $\lambda$ is for a rule to relax to a fixed point after a relatively short transient phase. As $\lambda$ is increased, rules tend to relax to periodic patterns, again after a relatively short transient phase. As $\lambda$ reaches a "critical value" $\lambda_c$, rules tend to have longer and longer transient phases. Additionally, the behavior in this regime exhibits long-lived, "complex"—non-periodic, but non-random—patterns. As $\lambda$ is increased further, the average transient length decreases, and rules tend to relax to apparently random space-time patterns. The actual value of $\lambda_c$ depends on $r$, $K$ and the actual path of CA found as $\lambda$ is incremented.

These four behavioral regimes roughly correspond to Wolfram's four classes. Langton's claim is that, as $\lambda$ is increased from 0 to $[1 - 1/K]$, the classes are passed through in the order 1, 2, 4, 3. He notes that as $\lambda$ is increased, "...one observes a *phase transition* between highly *ordered* and highly *disordered* dynamics, analogous to the phase transition between the *solid* and *fluid* states of matter." ([16], p. 13.)

According to Langton, as $\lambda$ is increased from $[1-1/K]$ to 1, the four regimes occur in the reverse order, subject to some constraints for $K > 2$ [16]. For two-state CA, since behavior is necessarily symmetric about $\lambda = 1/2$, there are two values of $\lambda_c$ at which the complex regime occurs.

---

[3]In [16] all states obeyed a "strong quiescence" requirement. For any state $s \in \{0, ..., K-1\}$, the neighborhood consisting entirely of state $s$ must map to $s$.

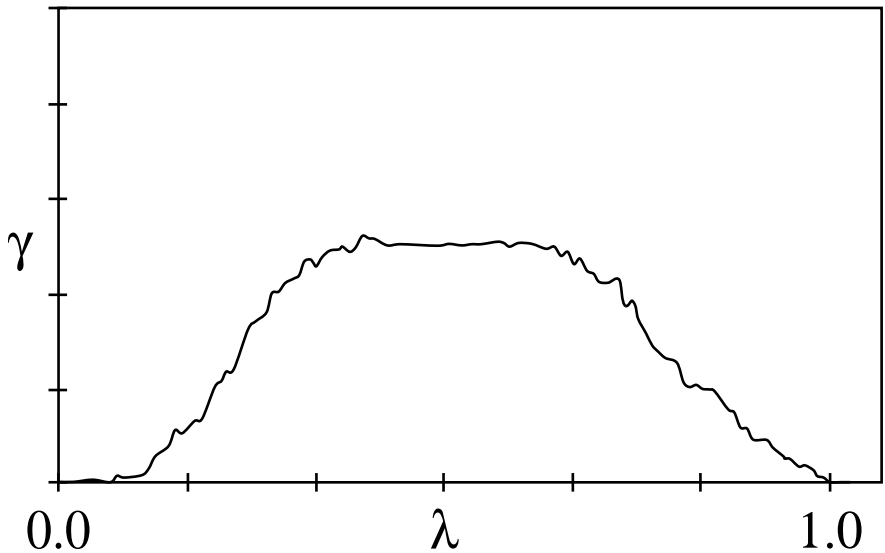

Figure 3: A graph of the average difference-pattern spreading rate $\gamma$ of a large number of randomly chosen $r = 3, K = 2$ CA, as a function of $\lambda$. Adapted from [23], with permission of the author. No vertical scale was provided there.

How is $\lambda_c$ determined? Following standard practice Langton used various statistics such as single-site entropy, two-site mutual information, and transient length to classify CA behavior. The additional step was to correlate behavior with $\lambda$ via these statistics. Langton's Monte Carlo samples showed there was some correlation between the statistics and $\lambda$. But the averaged statistics did not reveal a sharp transition in average behavior, a basic property of a phase transition in which macroscopic highly-averaged quantities do make marked changes. We note that Wootters and Langton gave evidence that in the limit of an increasing number of states the transition region narrows [32]. The main result indicates that in one class of two-dimensional infinite-state stochastic cellular automata there is a sharp transition in single-site entropy at $\lambda_c \approx 0.27$.

The existence of a critical $\lambda$ and the dependence of the critical region's width on $r$ and $K$ is less clear for finite-state CA. Nonetheless, Packard empirically determined rough values of $\lambda_c$ for $r = 3, K = 2$ CA by looking at the *difference-pattern spreading rate $\gamma$* as a function of $\lambda$ [23]. The spreading rate $\gamma$ is a measure of unpredictability in spatio-temporal patterns and so is one possible measure of chaotic behavior [21, 30]. It is analogous to, but not the same as, the Lyapunov exponent for continuous-state dynamical systems. In the case of CA it indicates the average propagation speed of information through space-time, though not the rate of production of local information.

At each $\lambda$ a large number of rules was sampled and for each CA $\gamma$ was estimated. The average $\gamma$ over the selected CA was taken as the average spreading rate at the given $\lambda$. The results are reproduced in Figure 3. As can be seen, at low and high $\lambda$'s, $\gamma$ vanishes; at intermediate $\lambda$ it is maximal, and in the "critical" $\lambda$ regions—centered about $\lambda \approx 0.25$ and $\lambda \approx 0.80$—it rises or falls gradually.

While not shown in Figure 3, for most $\lambda$ values $\gamma$'s variance is high. The same is true for single-site entropy and two-site mutual information as a function of $\lambda$ [16]. That is, the behavior of any *particular* rule at a given $\lambda$ might be very different from the *average* behavior at that value. Thus, the interpretations of these averages is somewhat problematic. This recounting of the behavioral structure of CA rule space as parameterized by $\lambda$ is based on statistics taken from Langton's and Packard's Monte Carlo simulations. Various problems in correlating $\lambda$ with behavior will be discussed in Section 8. A detailed analysis of some of these problems can be found in [2]. Other work on investigating the structure of CA rule space is reported in [18, 19].

The claim in [16] is that $\lambda$ predicts dynamical behavior well only when the space of rules is large enough. Apparently, $\lambda$ is not intended to be a good behavioral predictor for the space of elementary CA rules—$r = 1$, $K = 2$—and possibly $r = 3, K = 2$ rules as well.

## 5. CA Rule Space and Computation

Langton hypothesizes that a CA's computational capability is related to its average dynamical behavior, which $\lambda$ is claimed to predict [16]. In particular, he hypothesizes that CA capable of performing nontrivial computation—including universal computation—are most likely to be found in the vicinity of "phase transitions" between order and chaos, that is, near $\lambda_c$ values. The hypothesis relies on a basic observation of computation theory, that any form of computation requires memory—information storage—and communication—information transmission and interaction between stored and transmitted information. Above and beyond these properties, though, universal computation requires memory and communication over arbitrary distances in time and space. Thus complex computation requires significantly long transients and space-time correlation lengths; in the case of universal computation, arbitrarily long transients and correlations are required. Langton's claim is that these phenomena are most likely to be seen near $\lambda_c$ values—near "phase transitions" between order and chaos. This intuition is behind Langton's notion of "computation at the edge of chaos" for CA.[4]

## 6. Evolving CA

The empirical studies described above addressed only the relationship between $\lambda$ and the dynamical behavior of CA—as revealed by several statistics. Those studies did not correlate $\lambda$ or behavior with an independent measure of computation. Packard [23] addressed this issue by using a genetic algorithm (GA) [8, 12] to evolve CA rules to perform a particular computation. This experiment was meant to test two hypotheses: (1) CA rules able to perform complex computations are most likely to be found near $\lambda_c$ values; and (2) When CA rules are evolved to perform a complex computation, evolution will tend to select rules near $\lambda_c$ values.

### 6.1 The Computational Task and an Example CA

The original experiment consisted of evolving two-state—$s \in \{0, 1\}$—one-dimensional CA with $r = 3$. That is, the neighborhood of a cell consists of itself and its three neighbors on each side. The computational task for the CA is to decide whether or not the initial configuration contains more than half 1's. If the initial configuration contains more than half 1's, the desired behavior is for the CA, after some number of time steps, to relax to a fixed-point pattern of all 1's. If the initial configuration contains less than half 1's, the desired behavior is for the CA, after some number of time steps, to relax to a fixed-point pattern of all 0's. If the initial configuration contains exactly half 1's, then the desired behavior is undefined. This can be avoided in practice by requiring the CA lattice to be of odd length. Thus the desired CA has only two invariant patterns, either all 1's or all 0's. In the following we will denote the density of 1's in a lattice configuration by $\rho$, the density of 1's in the configuration at time $t$ by $\rho(t)$, and the threshold density for classification by $\rho_c$.

---

[4]This should be contrasted with the analysis of computation at the onset of chaos in [3, 4] and, in particular, with the discussion of the structure of CA space there.

Does the $\rho_c = 1/2$ classification task count as a "nontrivial" computation for a small-radius ($r \ll L$) CA? Though this term was not rigorously defined in [16] or [23], one possible definition might be any computation for which the memory requirement increases with $L$ (i.e., any computation which corresponds to the recognition of a non-regular language) and in which information must be transmitted over significant space-time distances (on the order of $L$). Under this definition the $\rho_c = 1/2$ classification task can be thought of as a nontrivial computation for a small radius CA. The effective minimum amount of memory is proportional to $log(L)$ since the equivalent of a counter register is required to track the excess of 1's in a serial scan of the initial pattern. And since the 1's can be distributed throughout the lattice, information transfer over long space-time distances must occur. This is supported in a CA by the non-local interactions among many different neighborhoods after some period of time.

Packard cited a $K = 2, r = 3$ rule constructed by Gacs, Kurdyumov, and Levin [5, 7], which purportedly performs this task. The Gacs-Kurdyumov-Levin (GKL) CA is defined by the following rule:

$$\text{If } s_i(t) = 0, \text{ then } s_i(t+1) = \text{majority } [s_i(t), s_{i-1}(t), s_{i-3}(t)];$$
$$\text{If } s_i(t) = 1, \text{ then } s_i(t+1) = \text{majority } [s_i(t), s_{i+1}(t), s_{i+3}(t)];$$

where $s_i(t)$ is the state of site $i$ at time $t$.

In words, this rule says that for each neighborhood of seven adjacent cells, if the state of the central cell is 0, then its new state is decided by a majority vote among itself, its left neighbor, and the cell two cells to the left away. Likewise, if the state of the central cell is 1, then its new state is decided by a majority vote among itself, its right neighbor, and the cell two cells to the right away.

Figure 2 gives space-time diagrams for the action of the GKL rule on an initial configuration with $\rho < \rho_c$ and on an initial configuration with $\rho > \rho_c$. It can be seen that, although the CA eventually converges to a fixed point, there is a transient phase during which a spatial and temporal transfer of information about local neighborhoods takes place, and this local information interacts with other local information to produce the desired final state. Very crudely, the GKL CA successively classifies "local" densities with the locality range increasing with time. In regions where there is some ambiguity, a "signal" is propagated. This is seen either as a checkerboard pattern propagated in both spatial directions or as a vertical white-to-black boundary. These signals indicate that the classification is to be made at a larger scale. Note that both signals locally have $\rho = \rho_c$; the result is that the signal patterns can propagate, since the density of patterns with $\rho = \rho_c$ is not increased or decreased under the rule. In a simple sense, this is the CA's "strategy" for performing the computational task.

It has been claimed that the GKL CA performs the $\rho_c = 1/2$ task [17], but actually this is true only to an approximation. The GKL rule was invented not for the purpose of performing any particular computational task, but rather as part of studies of reliable computation and phase transitions in one spatial dimension. The goal in the former, for example, was to find a CA whose behavior is robust to small errors in the rule's update of the configuration. It has been proved that the GKL rule has only two attracting patterns, either all 1's or all 0's [5]. Attracting patterns here are those invariant patterns which, when perturbed a small amount, return to the same pattern. It turns out that the basins of attraction for the all-1 and all-0 patterns are not precisely the initial configurations with $\rho > 1/2$ or $\rho < 1/2$, respectively.[5] On finite lattices the GKL rule does classify most initial configurations according to this criterion, but on a significant number the "incorrect" attractor is reached. One set of experimental measures of the GKL CA's classification performance is displayed in Figure 4. To make this plot, we ran the GKL CA on 500 randomly generated initial configurations at each of 19 densities $\rho \in [0.0, 1.0]$. The fraction of correct classifications was then plotted at each $\rho$. The rule was run either until a fixed point was reached or for a maximum number of time steps equal to $10 \times L$. This was done for CA with three different lattice sizes: $L \in \{149, 599, 999\}$.

Note that approximately 20% of the initial configurations with $\rho = \rho_c$ were misclassified. All the incorrect classifications are made for initial configurations with $\rho \approx \rho_c$. In fact, the worst performances occur at $\rho = \rho_c$. Interestingly, although the error region narrows with increasing lattice size, the performance at $\rho = \rho_c$ decreases when the lattice size is increased from 149 to 599.

The GKL rule table has $\lambda = 1/2$, not $\lambda = \lambda_c$. Since it appears to perform a computational task of some complexity, at a minimum it is a deviation from the "edge of chaos" hypothesis for CA computation. The GKL rule's $\lambda = 1/2$ puts it right at the center of the "chaotic" region in Figure 3. This may be puzzling, since clearly the GKL rule does not produce chaotic behavior during either its transient or asymptotic epochs—far from it, in fact. However, the $\lambda$ parameter was intended to correlate with "average" behavior of CA rules at a given $\lambda$ value. Recall that $\gamma$ in Figure 3 represent an *average* over a large number of randomly chosen CA rules and, while not shown in that plot, for most $\lambda$ values the variance in $\gamma$ is high. Thus, it can be claimed that the behavior of any *particular* rule at its $\lambda$ value might be very different from the *average* behavior at that value.

More to the point, though, we expect a $\lambda$ value close to 0.5 for a rule that performs well on the $\rho_c = 1/2$ task. This is largely because the task is symmetric with respect to the exchange of 1's and 0's. Suppose, for example, a rule that carries out the $\rho_c = 1/2$ task has $\lambda < 1/2$. This implies that there are more neighborhoods in the rule table that map to output bit 0 than to output bit 1. This, in turn, means that there will be *some* initial configurations with $\rho > \rho_c$ on which the action of the rule will *decrease* the number of 1's. And this is the opposite of the desired action. However, if the rule acts to *decrease* the number of 1's on an initial configuration with $\rho > \rho_c$, it risks producing an intermediate configuration with $\rho < \rho_c$, which then would lead (under the original assumption that the rule carries out the task correctly) to a fixed point of all 0's, misclassifying the initial configuration. A similar argument holds in the other direction if the rule's $\lambda$ value is greater than $1/2$. This informal argument shows that a rule with $\lambda \neq 1/2$ will misclassify certain initial configurations. Generally, the further away the rule is from $\lambda = 1/2$, the more such initial configurations there will be. Such rules may perform fairly well, classifying most initial configurations correctly. However, we expect any rule that performs reasonably well on this task—in the sense of being close to the GKL CA's average performance shown in Figure 4—to have a $\lambda$ value close to $1/2$.

---

[5]The terms "attractor" and "basin of attraction" are being used here in the sense of [5] and [11]. This differs substantially from the notion used in [33], for example. There "attractor" refers to any invariant or time-periodic pattern, and "basin of attraction" means that set of finite lattice configurations relaxing to it.

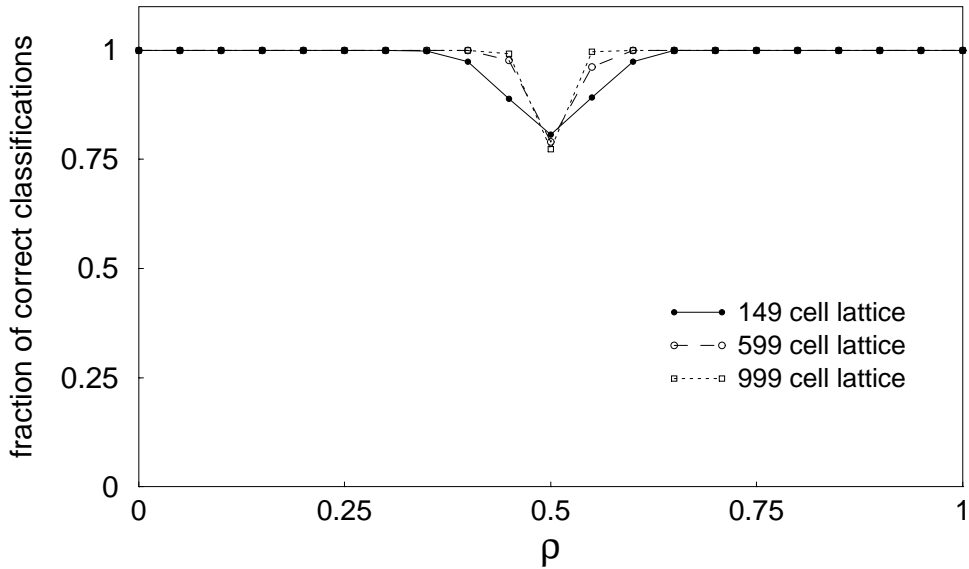

Figure 4: Experimental performance of the GKL rule as a function of $\rho(0)$ for the $\rho_c = 1/2$ task. Performance plots are given for three lattice sizes: $L = 149$ (the size of the lattice used in the GA runs), 599, and 999.

This analysis points to a problem with using this task as an evolutionary goal in order to study the relationship among evolution, computation, and $\lambda$. As was shown in Figure 3, for $r = 3, K = 2$ CA the $\lambda_c$ values occur at roughly 0.25 and 0.80, and one hypothesis that was to be tested by the original experiment is that the GA will tend to select rules close to these $\lambda_c$ values. But for the $\rho$-classification tasks, the range of $\lambda$ values required for good performance is simply a function of the task and, specifically, of $\rho_c$. For example, the underlying 0-1 exchange symmetry of the $\rho_c = 1/2$ task implies that if a CA exists to do the task at an acceptable performance level, then it has $\lambda \approx 1/2$. Even though this does not directly invalidate the adaptation hypothesis or claims about $\lambda$'s correlation with *average* behavior, it presents problems with using $\rho$-classification tasks as a way to gain evidence about a generic relation between $\lambda$ and computational capability.

## 6.2 The Original Experiment

Packard used a GA to evolve CA rules to perform the $\rho_c = 1/2$ task. His GA started out with a randomly generated initial population of CA rules. Each rule was represented as a bit string containing the output bits of the rule table. That is, the bit at position 0 in the string is the state to which the neighborhood 0000000 is mapped, the bit at position 1 in the string is the state to which the neighborhood 0000001 is mapped, and so on. The initial population was randomly generated but it was constrained to be uniformly distributed across $\lambda$ values between 0.0 and 1.0.

A given rule in the population was evaluated for ability to perform the classification task by choosing an initial configuration at random, running the CA on that initial configuration for some specified number of time steps, and at the final time step measuring the fraction of cells in the lattice that have the correct state. For initial configurations with $\rho > \rho_c$, the correct final state for each cell is 1, and for initial configurations with $\rho < \rho_c$, the correct final state for each cell is 0. For example, if the CA were run on an initial configuration with $\rho > \rho_c$ and at the final time step the lattice contained 90% 1's, the CA's score on that initial configuration would be 0.9.[6] The fitness of a rule was simply the rule's average score over a large number of initial configurations. For each rule in the population, Packard generated a large set of initial configurations that were uniformly distributed across $\rho$ values from 0 to 1.

Packard's GA worked as follows. At each generation:

1. The fitness of each rule in the population is calculated.
2. The population is ranked by fitness.
3. Some fraction of the lowest fitness rules are removed.
4. The removed rules are replaced by new rules formed by crossover and mutation from the remaining rules.

Crossover between two strings involves randomly selecting a position in the strings at random and exchanging parts of the strings before and after that position. Mutation involves flipping one or more bits in a string, with some low probability.

A diversity-enforcement scheme was also used to prevent the population from converging too early and losing diversity [22]. If a rule is formed that is too close in Hamming distance (i.e., the number of matching bits) to existing rules in the population, its fitness is decreased.

The results from Packard's experiment are displayed in Figure 5. The two histograms display the observed frequency of rules in the GA population as a function of $\lambda$, with rules merged from a number of different runs. The top graph gives this data for the initial generation. As can be seen, the rules are uniformly distributed over $\lambda$ values. The middle graph gives the same data for the final generation—in this case, after the GA has run for 100 generations. The rules now cluster around the two $\lambda_c$ regions, as can be seen by comparison with the difference-pattern spreading rate plot, reprinted here at the bottom of the figure. Note that each individual run produced rules at one or the other peak in the middle graph, so when the runs were merged together, both peaks appear [22]. Packard interpreted these results as evidence for the hypothesis that, when an ability for complex computation is required, evolution tends to select rules near the transition to chaos. He argues, like Langton, that this result intuitively makes sense because "rules near the transition to chaos have the capability to selectively communicate information with complex structures in space-time, thus enabling computation." ([23], p. 8).

---

[6]A slight variation on this method was used in [23]. Instead of measuring the fraction of correct states in the final lattice, the GA measured the fraction of correct states over configurations from a small number $n$ of final time steps [22]. This prevented the GA from evolving rules that were temporally periodic; viz. those with patterns that alternated between all 0's and all 1's. Such rules obtained higher than average fitness at early generations by often landing at the "correct" phase of the oscillation for a given initial configuration. That is, on the next time step the classification would have been incorrect. In our experiments we used a slightly different method to address this problem. This is explained in subsection 7.1.

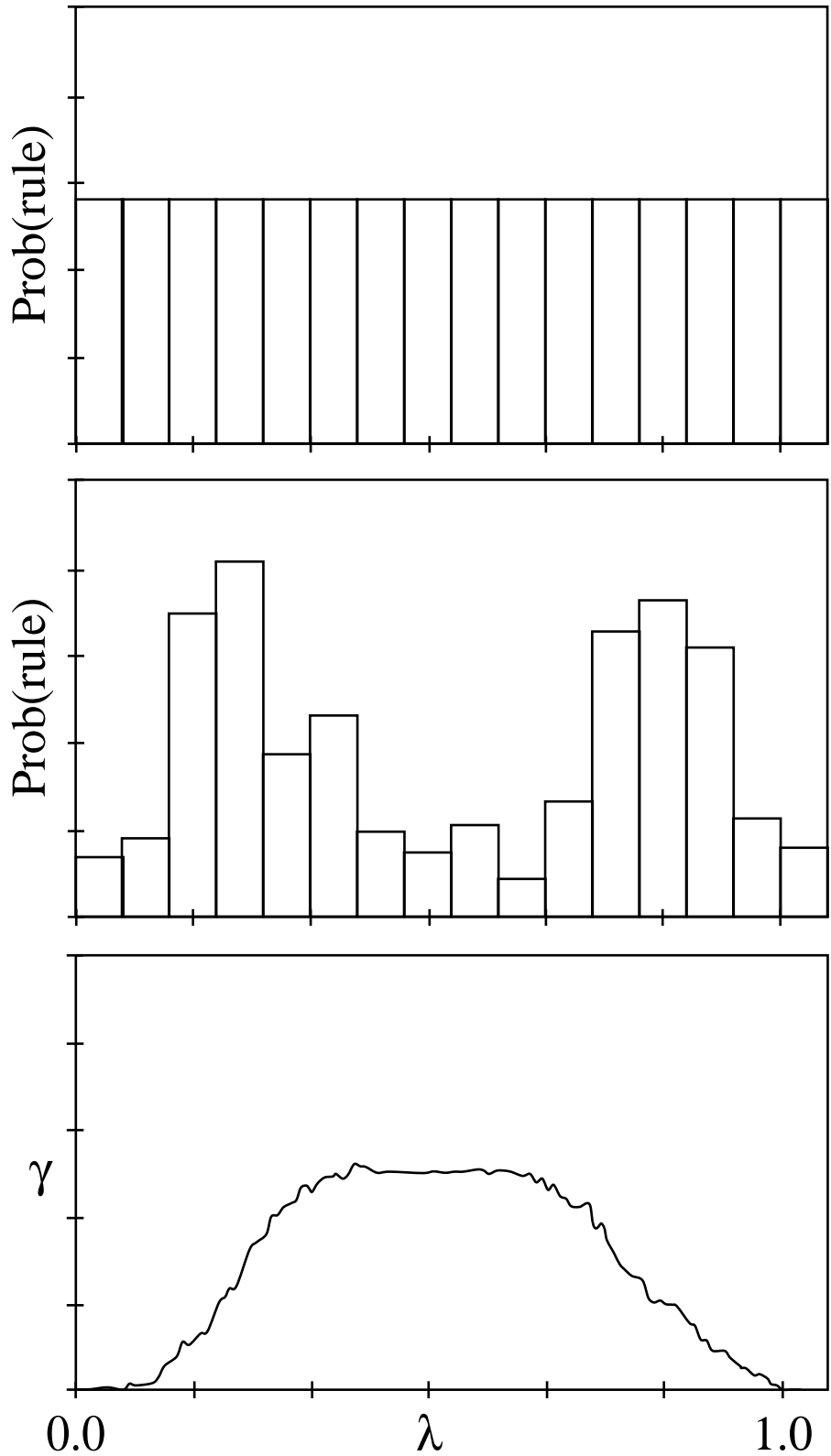

Figure 5: Results from the original experiment on GA evolution of CA for the $\rho_c = 1/2$ classification task. The top two figures are populations of CA at generations 0 and 100, respectively, versus $\lambda$. The bottom figure is Figure 3, reproduced here for reference. Adapted from [23], with permission of the author.

## 7. New Experiments

As the first step in a study of how well these general conclusions hold up, we carried out a set of experiments similar to that just described. We were unable to obtain some of the exact details of the original experiment's parameters, such as the exact population size for the GA, the mutation rate, and so on. As a result, we used what we felt were reasonable values for these various parameters. We carried out a number of parameter sensitivity tests which indicated that varying the parameters within small bounds did not change our qualitative results.

### 7.1 Details of Our Experiments

In our experiments, as in the original, the CA rules in the population all have $r = 3$ and $K = 2$. Thus the bit strings representing the rules are of length $2^{2r+1} = 128$ and the size of the search space is huge—the number of possible CA rules is $= 2^{128}$. The tests for each CA rule are carried out on lattices of length $L = 149$ with periodic boundary conditions. The population size is 100, which was roughly the population size used in the original experiment [22]. The initial population is generated at random, but constrained to be uniformly distributed among different $\lambda$ values. A rule's fitness is estimated by running the rule on 300 randomly generated initial configurations that are uniformly distributed over $\rho \in [0.0, 1.0]$. Exactly half the initial configurations have $\rho < \rho_c$ and exactly half have $\rho > \rho_c$.[7]

We allow each rule to run for a maximum number $M$ of iterations, where a new $M$ is selected for each rule from a Poisson distribution with mean 320. This is the measured maximum amount of time for the GKL CA to reach an invariant pattern over a large number of initial configurations on lattice size 149.[8] A rule's fitness is its average score—the fraction of cell states correct at the last iteration—over the 300 initial configurations. We term this fitness function *proportional fitness* to contrast with a second fitness function—*performance fitness*—which will be described below. A new set of 300 initial configurations is generated every generation. At each generation, all the rules in the population are tested on this set. Notice that this fitness function is stochastic—the fitness of a given rule may vary from generation to generation depending on the set of 300 initial configurations used in testing it.

---
[7]It was necessary to have this exact symmetry in the initial configurations at each generation to avoid early biases in the $\lambda$ of selected rules. If, say, 49% of the initial configurations have $\rho < \rho_c$ and 51% of initial configurations have $\rho > \rho_c$, high $\lambda$ rules would obtain slightly higher fitness than low $\lambda$ rules since high $\lambda$ rules will map most initial configurations to all 1's. A rule with, say, $\lambda \approx 1$ would in this case classify 51% of the initial configurations correctly whereas a rule with $\lambda \approx 0$ would classify only 49% correctly. But such slight differences in fitness have a large effect in the initial generation, when all rules have fitness close to 0.5, since the GA selects the 50 *best* rules, even if they are only very slightly better than the 50 *worst* rules. This biases the representative rules in the early population. And this bias can persist well into the later generations.

Our GA is similar to Packard's. In our GA, the fraction of new strings in the next generation—the "generation gap"—is 0.5. That is, once the population is ordered according to fitness, the top half of the population, the set of "elite" strings, is copied without modification into the next generation. For GA practitioners more familiar with nonoverlapping generations, this may sound like a small generation gap. However, since testing a rule on 300 "training cases" does not necessarily provide a very reliable gauge of what the fitness would be over a larger set of training cases, our selected gap is a good way of making a "first cut" and allowing rules that survive to be tested over more initial configurations. Since a new set of initial configurations is produced every generation, rules that are copied without modification are always retested on this new set. If a rule performs well and thus survives over a large number of generations, then it is likely to be a genuinely better rule than those that are not selected, since it has been tested with a large set of initial configurations. An alternative method would be to test every rule in every generation on a much larger set of initial configurations, but given the amount of compute time involved, that method seems unnecessarily wasteful. Much too much effort, for example, would go into testing very weak rules, which can safely be weeded out early using our method.

The remaining half of the population for each new generation is created by crossover and mutation from the previous generation's population.[9] Fifty pairs of parent rules are chosen at random with replacement from the entire previous population. For each pair, a single crossover point is selected at random, and two offspring are created by exchanging the subparts of each parent before and after the crossover point. The two offspring then undergo mutation. A mutation consists of flipping a randomly chosen bit in the string. The number of mutations for a given string is chosen from a Poisson distribution with a mean of 3.8 (this is equivalent to a per-bit mutation rate of 0.03). Again, to GA practitioners this may seem to be a high mutation rate, but one must take into account that at every generation, half the population is being copied without modification.

## 7.2 Results of Proportional-Fitness Experiment

---

[8] It may not be necessary to allow the maximum number of iterations $M$ to vary. In some early tests with smaller sets of fixed initial configurations, though, we found the same problem Packard reported [22]: that if $M$ is fixed, then period-2 rules evolve that alternate between all 0's and all 1's. These rules adapted to the small set of initial configurations and the fixed $M$ by landing at the "correct" pattern for a given initial configuration at time step $M$, only to move to the opposite pattern and so wrong classification at time step $M + 1$. These rules did very poorly when tested on a different set of initial configurations—evidence for "over-fitting".

[9] This method of producing the non-elite strings differs from that in [23], where the non-elite strings were formed from crossover and mutation among the elite strings only rather than from the entire population. We observed no statistically significant differences in our tests using the latter mechanism other than a modest difference in time scale.

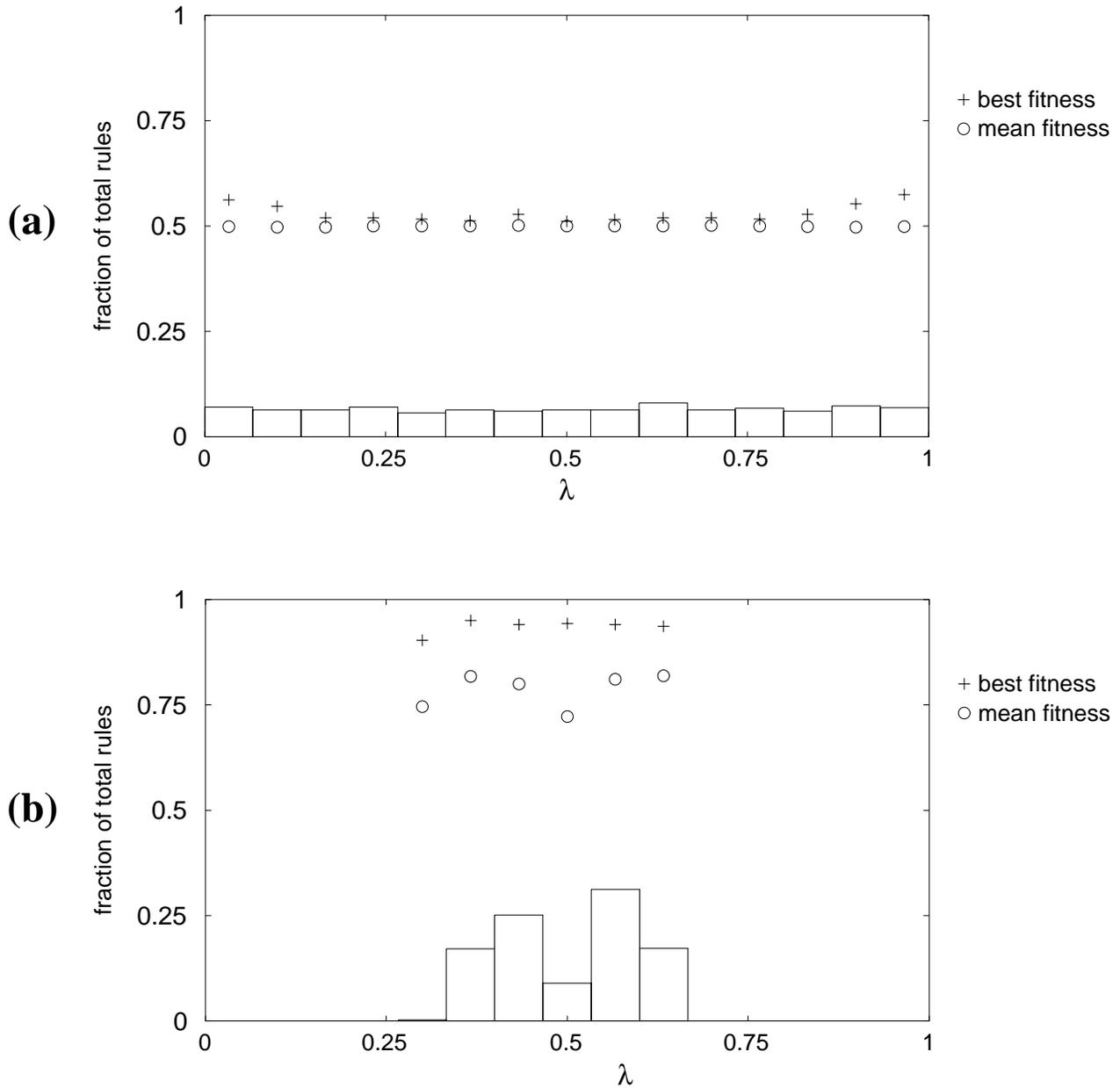

Figure 6: Results from our experiment with proportional fitness. The top histogram (a) plots as a function of $\lambda$ the frequencies of rules merged from the initial generations of 30 runs. The bottom histogram (b) plots the frequencies of rules merged from the final generations (generation 100) of these 30 runs. Following [23] the $x$-axis is divided into 15 bins of length 0.0667 each. The rules with $\lambda = 1.0$ are included in the rightmost bin. In each histogram the best (cross) and mean (circle) fitnesses are plotted for each bin. (The $y$-axis interval for fitnesses is also [0,1]).

We performed 30 different runs of the GA with the parameters described above, each with a different random-number seed. On each run the GA was iterated for 100 generations. We found that running the GA for longer than this, up to 300 generations, did not result in improved fitness. The results of this set of runs are displayed in Figure 6. Figure 6(a) is a histogram of the frequency of rules in the initial populations as a function of $\lambda$, merging together the rules from all 30 initial populations; thus the total number of rules represented in this histogram is 3000. The $\lambda$ bins in this histogram are the same ones that were used by Packard, each of width 0.0667. Packard's highest bin contained only rules with $\lambda = 1$, that is, rules that consist of all 1's. We have merged this bin with the immediately lower bin.

As was said earlier, the initial population consists of randomly generated rules uniformly spread over the $\lambda$ values between 0.0 and 1.0. Also plotted are the mean and best fitness values for each bin. These are all around 0.5, which is expected for a set of randomly generated rules under this fitness function. The best fitnesses are slightly higher in the very low and very high $\lambda$ bins. This is because rules with output bits that are almost all 0's (or 1's) correctly classify all low density (or all high density) initial configurations. In addition these CA obtain small partial credit on some high density (low density) initial configurations. Such rules thus have fitness sightly higher than 0.5.

Figure 6(b) shows the histogram for the final generation (100), merging together rules from the final generations of all 30 runs. Again the mean and best fitness values for each bin are plotted.

In the final generation the mean fitnesses in each bin are all around 0.8. The exceptions are the central bin with a mean fitness of 0.72 and the leftmost bin with a mean fitness of 0.75. The leftmost bin contains only five rules—each at $\lambda \approx 0.33$, right next to the the bin's upper $\lambda$ limit. The standard deviations of fitness for each bin, not shown in the figure, are all approximately 0.15, except the leftmost bin, which has a standard deviation of 0.20. The best fitnesses for each bin are all between 0.93 and 0.95, except the leftmost bin which has a best fitness of 0.90. Under this fitness function the GKL rule has fitness $\approx 0.98$; the GA never found a rule with fitness above 0.95.

As was mentioned above, the fitness function is stochastic: a given rule might be assigned a different fitness each time the fitness function is evaluated. The standard deviation under the present fitness scheme on a given rule is approximately 0.015. This indicates that the differences among the best fitnesses plotted in the histogram are not significant, except for that in the leftmost bin.

The lower mean fitness in the central bin is due to the fact that the rules in that bin largely come from non-elite rules generated by crossover and mutation in the final generation. This is a combinatorial effect: the density of CA rules as a function of $\lambda$ is very highly peaked about $\lambda = 1/2$, as already noted. We will return to this "combinatorial drift" effect shortly. Many of the rules in the middle bin have not yet undergone selection and thus tend to have lower fitnesses than rules that have been selected in the elite. This effect disappears in Figure 7, which includes only the elite rules at generation 100 for the 30 runs. As can be seen, the difference in mean fitness disappears and the height of the central bin is decreased by half.

The results presented in Figure 6(b) are strikingly different from the results of the original experiment. In the final generation histogram in Figure 5, most of the rules clustered around either $\lambda \approx 0.24$ or $\lambda \approx 0.83$. Here, though, there are no rules in these $\lambda_c$ regions. Rather, the rules cluster much closer—with a ratio of variances of 4 between the two distributions—to $\lambda \approx 0.5$. Recall this clustering is what we expect from the basic 0-1 exchange symmetry of the $\rho_c = 1/2$ task.

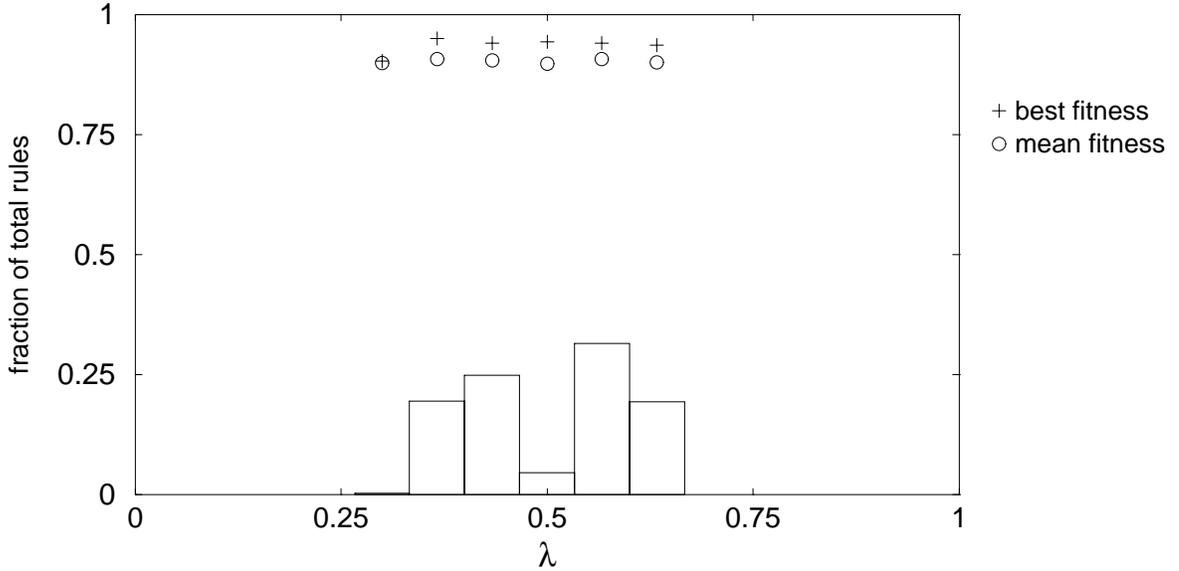

Figure 7: Histogram including only the elite rules from the final generations of the 30 runs (cf. Figure 6(b)) with the proportional-fitness function.

One rough similarity is the presence of two peaks centered around a dip at $\lambda \approx 0.5$—a phenomenon which we will explain shortly and which is a key to understanding how the GA is working. But there are significant differences, even within this similarity. In the original experiments the peaks are in bins centered about $\lambda \approx 0.24$ and $\lambda \approx 0.83$. In Figure 6(b), though, the peaks are very close to $\lambda = 1/2$, being centered in the neighboring bins—those with $\lambda \approx 0.43$ and $\lambda \approx 0.57$. Thus, the ratio of original to current peak spread is roughly a factor of 4. Additionally, in the final-generation histogram of Figure 5 the two highest bin populations are roughly five times as high as the central bin, whereas in the final-generation histogram of Figure 6(b) the two highest bins are roughly three times as high as the central bin. Finally, the final-generation histogram in Figure 5 shows the presence of rules in every bin, but in the new final-generation histogram, there are only rules in six of the central bins.

Similar to the original experiment, we found that on any given run the population was clustered about one or the other peak but not both. Thus, in the histograms that merge all runs, two peaks appear. This is illustrated in Figure 8, which displays histograms from the final generation of two individual runs. In one of these runs the population clustered to the left of the central bin, in the other run it clustered to the right of the center. The fact that different runs result in different clustering locations is why we performed many runs and merged the results rather than performing a single run with a much larger population. The latter method might have yielded only one peak. Said a different way, independent of the population size a given run will be driven by and the population organized around the fit individuals that appear earliest. Thus, examining an ensemble of individual runs reveals more details of the evolutionary dynamics.

The asymmetry in the heights of the two peaks in Figure 6(b) results from a small statistical asymmetry in the results of the 30 runs. There were 14 out of 30 runs in which the rules clustered at the lower $\lambda$ bin and 16 out of 30 runs in which the rules clustered at the higher $\lambda$ bin. This difference is not significant, but explains the small asymmetry in the peaks' heights.

We extended 16 of the 30 runs to 300 generations, and found that not only do the fitnesses not increase further, but the basic shape of the histogram does not change significantly.

## 7.3 Effects of Drift

The results of our experiments suggest that, for the $\rho_c = 1/2$ task, an evolutionary process modeled by a genetic algorithm tends to select rules with $\lambda \approx 1/2$. This is what we expect from the theoretical discussion given above concerning this task and its symmetries. We will delay until the next section a discussion of the curious feature near $\lambda = 1/2$, viz. the dip surrounded by two peaks. Instead, here we focus on the larger-scale clustering in that $\lambda$ region.

To understand this clustering we need to understand the degree to which the selection of rules close to $\lambda = 1/2$ is due to an intrinsic selection pressure and the degree to which it is due to "drift". By "drift" we refer to the force that derives from the combinatorial aspects of CA space as explored by random selection ("genetic drift") along with the effects of crossover and mutation. The intrinsic effects of random selection with crossover and mutation are to move the population, irrespective of any selection pressure, to $\lambda = 1/2$. This is illustrated by the histogram mosaic in Figure 9. These histograms show the frequencies of the rules in the population as a function of $\lambda$ every 5 generations, from 30 runs on which selection according to fitness was turned off. That is, on these runs, the fitness of the rules in the population was never calculated, and at each generation the selection of the elite group of strings was performed at random. Everything else about the runs remains the same as before. Since there is no selection, drift is the only force at work here. As can be seen, under the effects of random selection, crossover, and mutation, by generation 10 the population has largely drifted to the region of $\lambda = 1/2$ and this clustering becomes increasingly pronounced as the run continues.

This drift to $\lambda = 1/2$ is related to the combinatorics of the space of bit strings. For binary CA rules with neighborhood size $N$ ($= 2r + 1$), the space consists of all $2^{2^N}$ binary strings of length $2^N$. Denoting the subspace of CA with a fixed $\lambda$ and $N$ as $\mathrm{CA}(\lambda, N)$, we see that the size of the subspace is binomially distributed with respect to $\lambda$:

$$|\mathrm{CA}(\lambda, N)| = \binom{2^N}{\lambda 2^N}.$$

The distribution is symmetric in $\lambda$ and tightly peaked about $\lambda = 1/2$ with variance $\propto 2^{-N}$. Thus, the vast majority of rules is found at $\lambda = 1/2$. The steepness of the binomial distribution near its maximum gives an indication of the magnitude of the drift "force". Note that the last histogram in Figure 9 gives the GA's rough approximation of this distribution.

Drift is thus a powerful force moving the population to cluster around $\lambda = 1/2$. For comparison, Figure 10 gives the rule-frequency-versus-$\lambda$ histograms for the 30 runs of our proportional-fitness experiment every five generations. The last histogram in this figure is the same one that was displayed in Figure 6(b). (Figure 10 gives the merged data from the entire population of each run every five generations. A similar mosaic plotting only the elite strings at each generation looks qualitatively similar.)

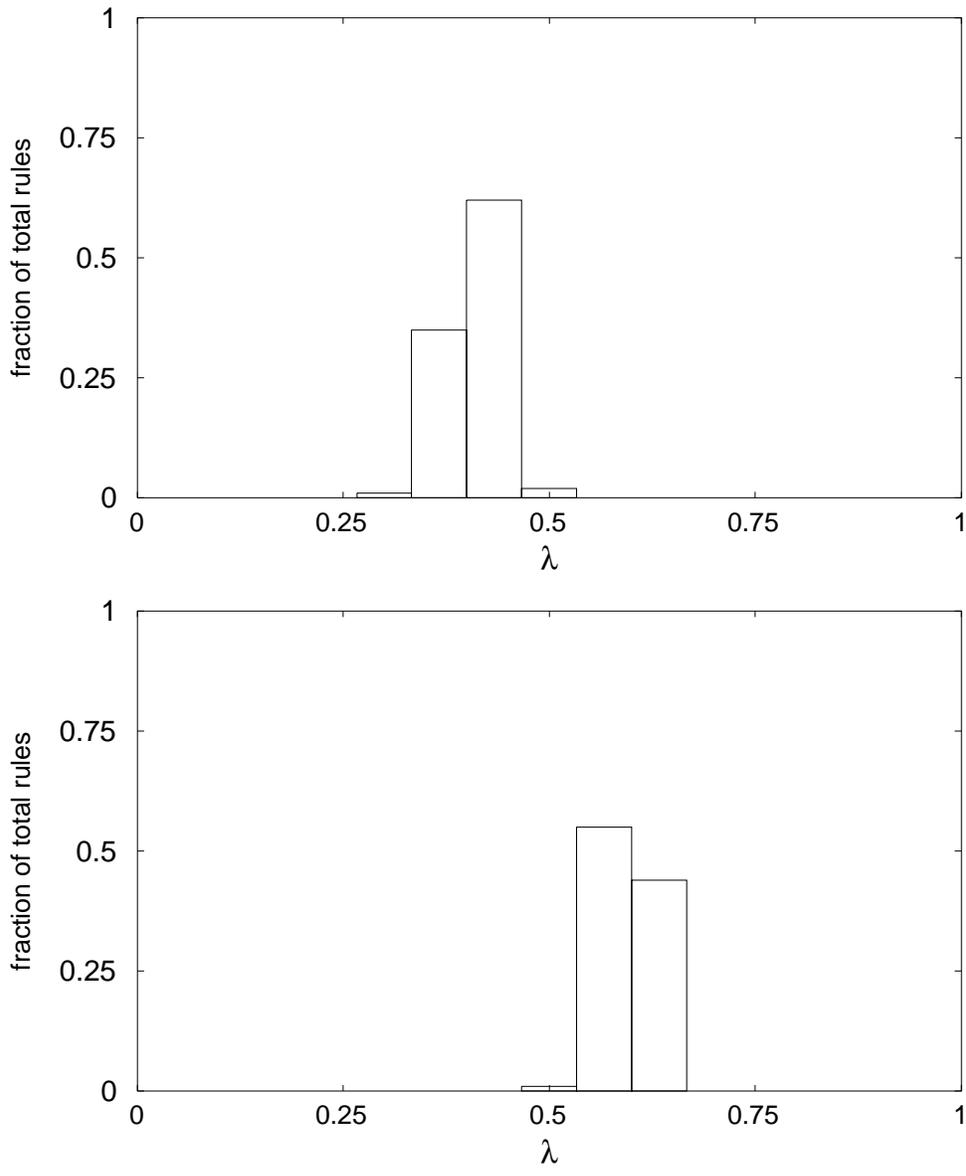

Figure 8: Histograms from the final generations of two individual runs of the GA employing proportional fitness. Each run had a population of 100 rules. The final distribution of rules in each of the 30 runs we performed resembled one or the other of these two histograms.

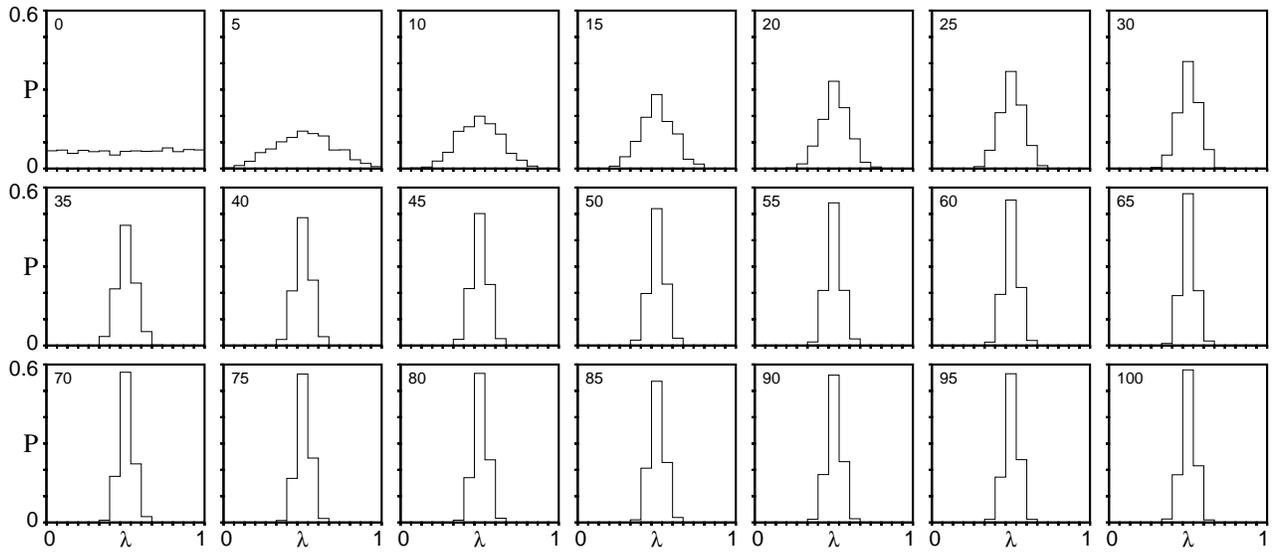

Figure 9: *No-selection mosaic:* Rule-frequency-versus-$\lambda$ histograms given every five generations for populations evolved under the genetic algorithm with no selection; that is, the fitness function was not calculated. As before, each histogram is merged from 30 runs; each run had a population of 100 rules. The generation number is given in the upper left corner of each histogram.

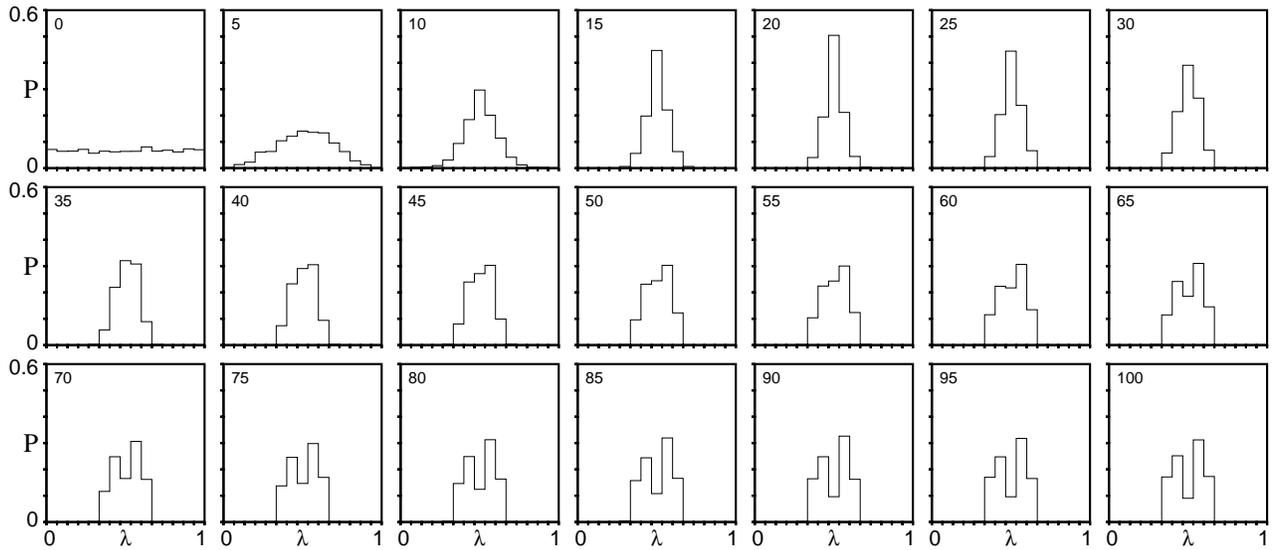

Figure 10: *Proportional-fitness mosaic:* Rule-frequency-versus-$\lambda$ histograms given every five generations, merged from the 30 GA runs with proportional fitness. Each run had a population of 100 rules.

Figure 10 looks very similar to Figure 9 up to generation 35. The main difference in generations 0–30 is that Figure 10 indicates a more rapid peaking about $\lambda = 1/2$. The increased speed of movement to the center over that seen in Figure 9 is presumably due to the additional evolutionary pressure of proportional fitness. At generation 35, something new appears. The peak in the center has begun to shrink significantly and the two surrounding bins are beginning to rival it in magnitude. By generation 40 the right-of-center bin has exceeded the central bin, and by generation 65 the histogram has developed two peaks surrounding a dip in the center. The dip becomes increasingly pronounced as the run continues, but stabilizes by generation 85 or so.

The differences between Figure 10 and Figure 9 over all 100 generations shows that the population's structure in each generation is not entirely due to drift. Indeed, after generation 35 the distinctive features of the population indicates new, qualitatively different, and unique properties due to the selection mechanism. The two peaks represent a symmetry breaking in the evolutionary process—the rules in each individual run initially are clustered around $\lambda = 1/2$ but move to one side or the other of the central bin by around generation 35. The causes of this symmetry breaking will be discussed in the next subsection.

### 7.4 Evolutionary Mechanisms: Symmetry Breaking and the Dip at $\lambda = 1/2$

At this point we move away from questions related to the original experiment and instead concentrate on the mechanisms involved in producing our results. Two major questions need to be answered: Why in the final generation are there significantly fewer rules in the central bin than in the two surrounding bins? And what causes the symmetry breaking that begins near generation 35 seen in Figure 10?

In the briefest terms, the answer, obtained by detailed analysis of the 30 GA runs, is the following. The course of CA evolution under our GA roughly falls into four "strategy" epochs. Each epoch is associated with an innovation discovered by the GA for solving the problem. Though the absolute time at which these innovations appear in each run varies somewhat, each run basically passes through each of these four epochs in succession. The epochs are shown in Figure 11, which plots the best fitness, the mean fitness of the elite strings, and the mean fitness of the population versus generation for one typical run of the GA. The beginnings of epochs 2 through 4 are pointed out on the best-fitness plot. Epoch 1 begins at generation 0.

#### Epoch 1: Randomly generated rules

The first epoch starts at generation 0, when the best fitness in the initial generation is approximately 0.5, and the $\lambda$ values are uniformly distributed between 0.0 and 1.0. No rule is much fitter than any other rule, though as was seen in Figure 6(a), rules with very low and very high $\lambda$ tend to have slightly higher fitness. The strategy here—if it can be called this at all—derives from only the most elementary aspect of the task. Rules either specialize for $\rho > \rho_c$ configurations by mapping high-density neighborhoods in the CA rule table to 1 or specialize for $\rho < \rho_c$ configurations by mapping low-density neighborhoods to 0.

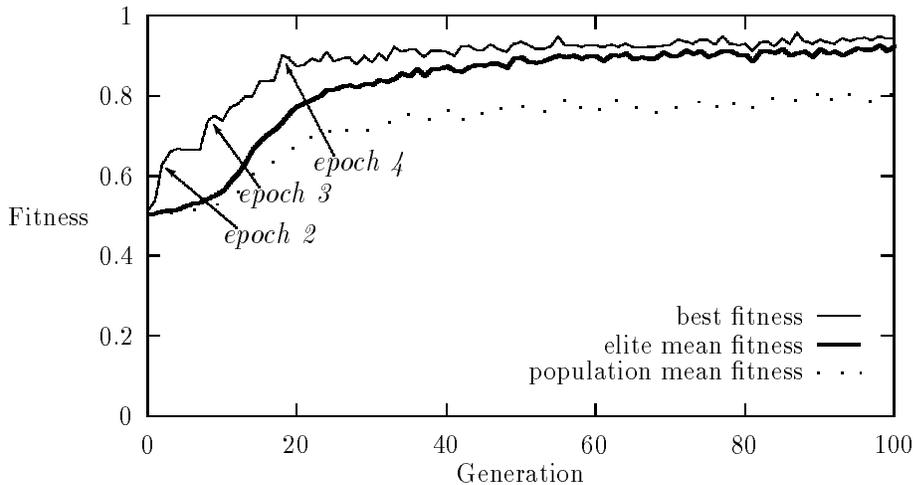

Figure 11: Best fitness, elite mean fitness, and population mean fitness versus generation for one typical run. The beginnings of epochs 2–4 are pointed out on the best-fitness plot. Epoch 1 begins at generation 0.

**Epoch 2: Discovery of two halves of the rule table**

The second epoch begins when a rule is discovered in which most neighborhood patterns in the rule table that have $\rho < \rho_c$ map to 0 and most neighborhood patterns in the rule table that have $\rho > \rho_c$ map to 1. This is roughly correlated with the left and right halves of the rule table: namely, neighborhoods 0000000 to 0111111 and 1000000 to 1111111, respectively. Such a strategy is presumably easy for the GA to discover due to single-point crossover's tendency to preserve contiguous sections of the rule table. It differs from the accidental strategy of epoch 0 in that there is now an organization to the rule table: output bits are roughly associated with densities of neighborhood patterns. It is the first significant attempt at distinguishing initial configurations with more 1's than 0's and vice versa. Under our fitness function, the fitness of such rules is approximately between 0.6 and 0.7, which is significantly higher than the fitness of the initial random rules. This innovation typically occurs between generations 1 and 10; in the run displayed in Figure 11 it occurred in generation 2, and can be seen as the steep rise in the best-fitness plot at that generation. All such rules tend to have $\lambda$ close to 0.5. There are many possible variations on these rules with similar fitness, so such rules—all close to $\lambda = 1/2$—begin to dominate in the population. This, along with the natural tendency for the population to drift to $\lambda = 1/2$, is the cause of the clustering around $\lambda = 1/2$ seen by generation 10 in Figure 10. For the next several generations the population tends to explore small variations on this broad strategy. This can be seen in Figure 11 as the leveling off in the best-fitness plot between generations 2 and 10.

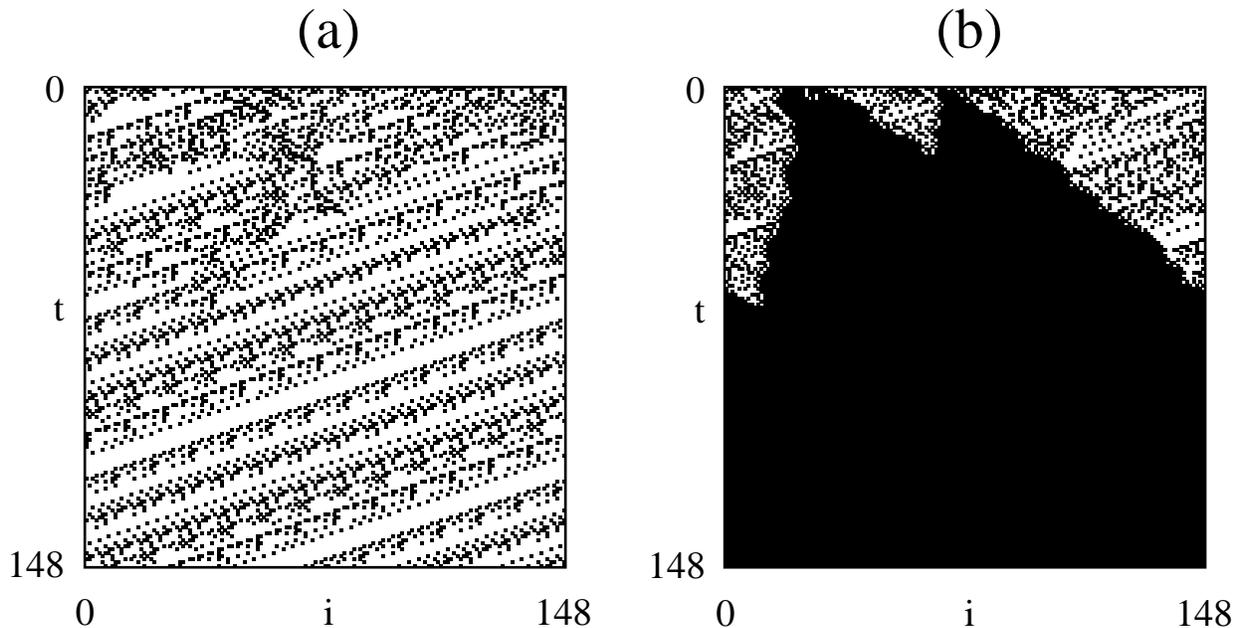

Figure 12: Space-time diagrams of one epoch-3 rule with $\lambda \approx 0.41$ that increases sufficiently large blocks of adjacent or nearly adjacent 1's. Both diagrams have $L = 149$ and are iterated for 149 time steps (the time displayed here is shorter than the actual time allotted under the GA). In (a) $\rho(0) \approx 0.40$ and $\rho(148) \approx 0.17$. In (b) $\rho(0) \approx 0.54$ and $\rho(148) = 1.0$. Thus, in (a) the classification is incorrect, but partial credit is given; in (b) it is correct.

**Epoch 3: Growing blocks of 1's or 0's**

The next epoch begins when the GA discovers one of two new strategies. The first strategy is to increase the size of a sufficiently large block of adjacent or nearly adjacent 0's; the second strategy is to increase the size of a sufficiently large block of adjacent or nearly adjacent 1's.

Examples of these two strategies are illustrated in Figures 12 and 13. These figures give space-time diagrams from two rules that marked the beginning of this epoch in two different runs of the GA. Figure 12 illustrates the action of a rule discovered at generation 9 of one run. This rule has $\lambda \approx 0.41$, which means that the rule maps most neighborhoods to 0. Its strategy is to map initial configurations to mostly 0's—the configurations it produces have $\rho < \rho_c$, unless the initial configuration contains a sufficiently large block of 1's, in which case it increases the size of that block. The left space-time diagram Figure 12(a) shows how the rule evolves an initial configuration with $\rho < \rho_c$, to a final lattice with mostly 0's. This produces a fairly good score. The right space-time diagram Figure 12(b) shows how the rule evolves an initial configuration with $\rho > \rho_c$. The initial configuration contains a few sufficiently large blocks of adjacent or nearly adjacent 1's, and the size of these blocks is quickly increased to yield a final lattice with all 1's for a perfect score. The fitness of this rule at generation 9 was $\approx 0.80$.

Figure 13 illustrates the action of a second rule, discovered at generation 20 in another run. This rule has $\lambda \approx 0.58$, which means that the rule maps most neighborhoods to 1. Its strategy is the inverse of the previous rule. It maps initial configurations to mostly 1's unless the initial configuration contains a sufficiently large block of 0's, in which case it increases the size of that block. The left space-time diagram (a) illustrates this for an initial configuration with $\rho < \rho_c$; here a sufficiently large block of 0's appears in the initial configuration and is increased in size, yielding a perfect score. The right space-time diagram (b) shows the action of the same rule on an initial configuration with $\rho > \rho_c$. Most neighborhoods are mapped to 1 so the final configuration contains mostly 1's, yielding a fairly high score. The fitness of this rule at generation 20 was $\approx 0.87$.

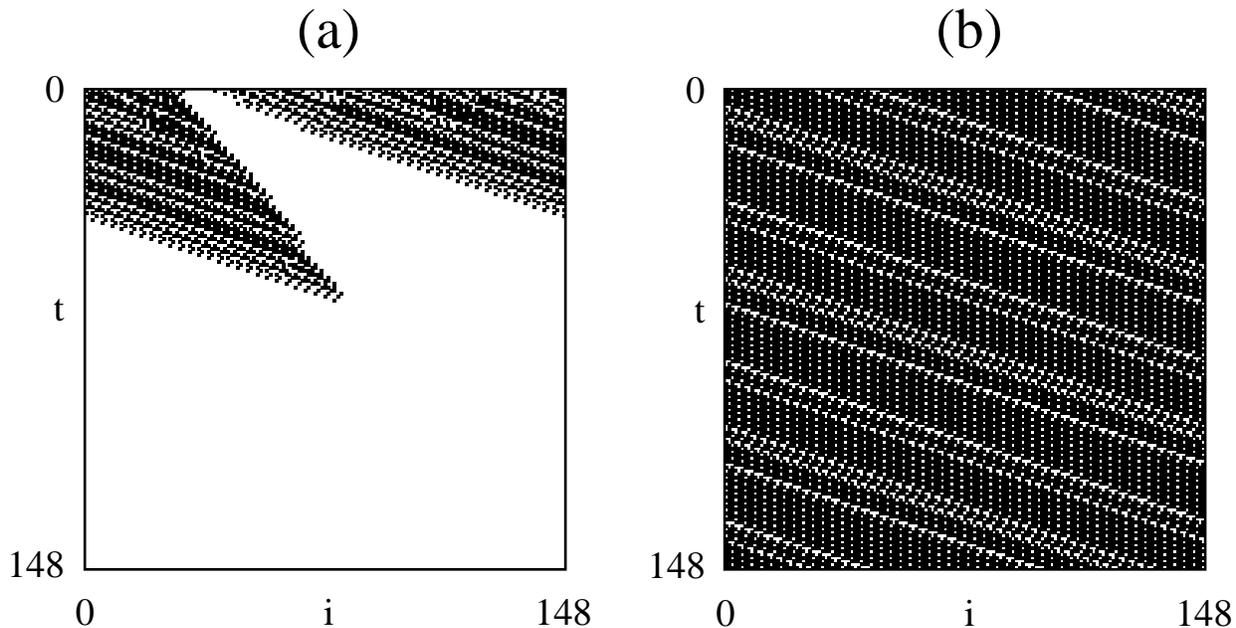

Figure 13: Space-time diagrams of one epoch-3 rule with $\lambda \approx 0.58$ that increases sufficiently large blocks of adjacent or nearly adjacent 0's. In (a) the initial configuration with $\rho \approx 0.42$ maps to a correct classification pattern of all 0's. In (b) the initial configuration with $\rho \approx 0.56$ is not correctly classified ($\rho(148) \approx 0.75$) but partial credit is given.

The general idea behind these two strategies is to rely on statistical fluctuations in the initial configurations. An initial configuration with $\rho > \rho_c$ is likely to contain a sufficiently large block of adjacent or nearly adjacent 1's. The rule then increases this region's size to yield the correct classification. Similarly, this holds for the CA in Figure 13 with respect to blocks of 0's in initial configurations with $\rho < \rho_c$. In short, these strategies are assuming that the presence of a sufficiently large block of 1's or 0's is a good predictor of $\rho(0)$.

Similar strategies were discovered in every run. They typically emerge by generation 20. A given strategy either increased blocks of 0's or blocks of 1's, but not both. These strategies result in a significant jump in fitness: typical fitnesses for the first instances of such strategies range from 0.75 to 0.85. This jump in fitness can be seen in the run of Figure 11 at approximately generation 10, and is marked as the beginning of epoch 3. This is the first epoch in which a substantial increase in fitness is associated with a symmetry breaking in the population, which will be explained below.

The first instances of epoch-3 strategies typically have a number of problems. As can be seen in Figures 12 and 13, the rules often rely on partial credit to achieve fairly high fitness on structurally incorrect classification. They typically do not get perfect scores on many initial configurations. The rules also often make mistakes in classification. Three common types of classification errors are illustrated in Figure 14. Figure 14(a) illustrates a rule increasing a too-small block of 1's and thus misclassifying an initial configuration with $\rho < \rho_c$. Figure 14(b) illustrates a rule that does not increase blocks of 1's fast enough on an initial configuration with $\rho > \rho_c$, leaving many incorrect bits in the final lattice. Figure 14(c) illustrates the *creation* of a block of 1's that did not appear in an initial configuration with $\rho < \rho_c$, ultimately leading to a misclassification. The rules that produced these diagrams come from epoch 3 in various GA runs.

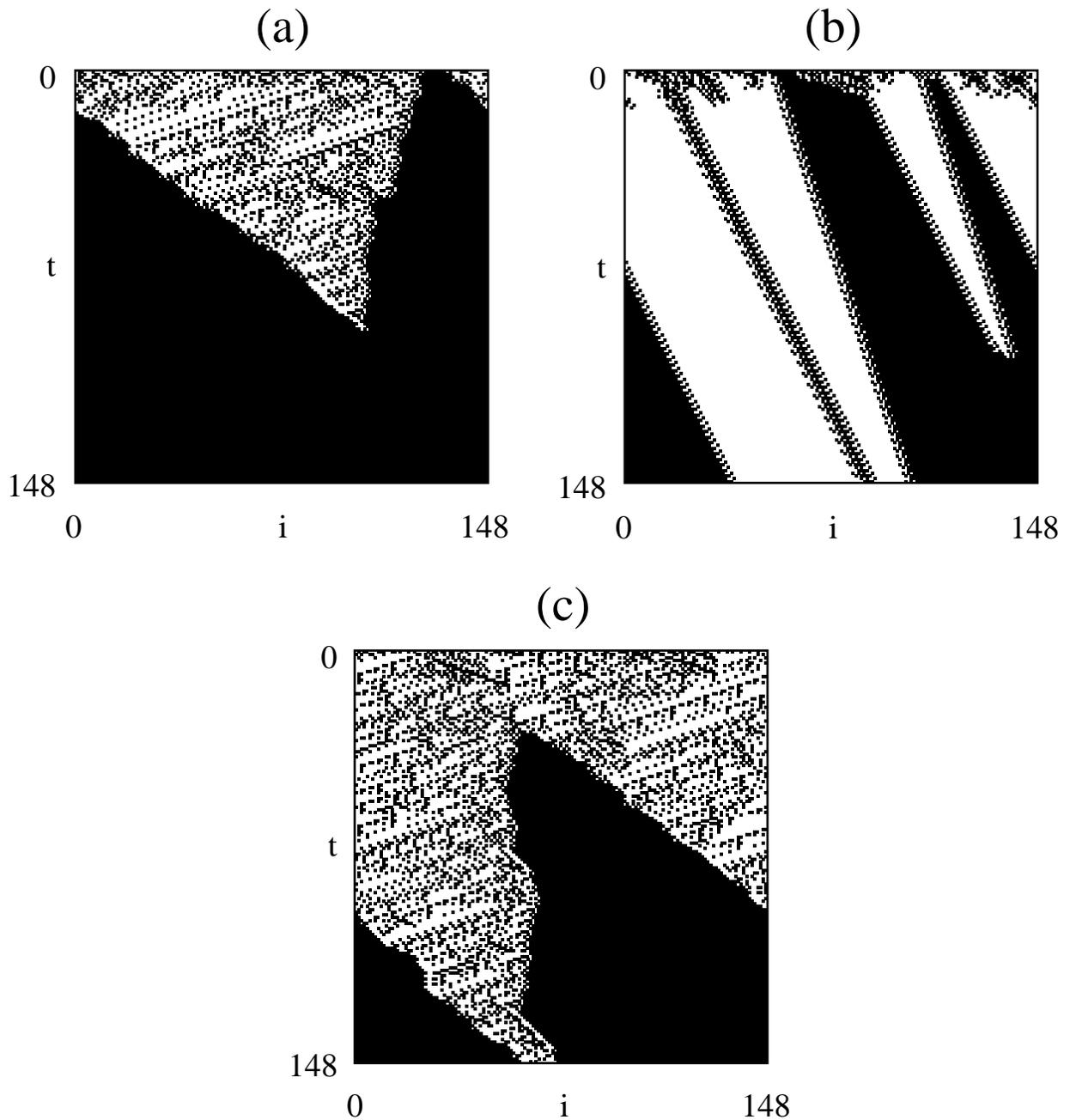

Figure 14: Space-time diagrams illustrating three types of classification errors committed by epoch-3 rules: (a) growing a block of 1s in a sea of $\rho < \rho_c$, (b) growing a block of 1's for an initial configuration with $\rho > \rho_c$ too slowly (the correct fixed point of all 1's does not occur until iteration 480), and (c) generating a block of 1's from a sea of $\rho < \rho_c$ and growing it so that $\rho > \rho_c$ (the incorrect fixed point of all 1's occurs at iteration 180). The initial configuration densities are (a) $\rho(0) \approx 0.39$, (b) $\rho(0) \approx 0.59$, and (c) $\rho(0) \approx 0.45$.

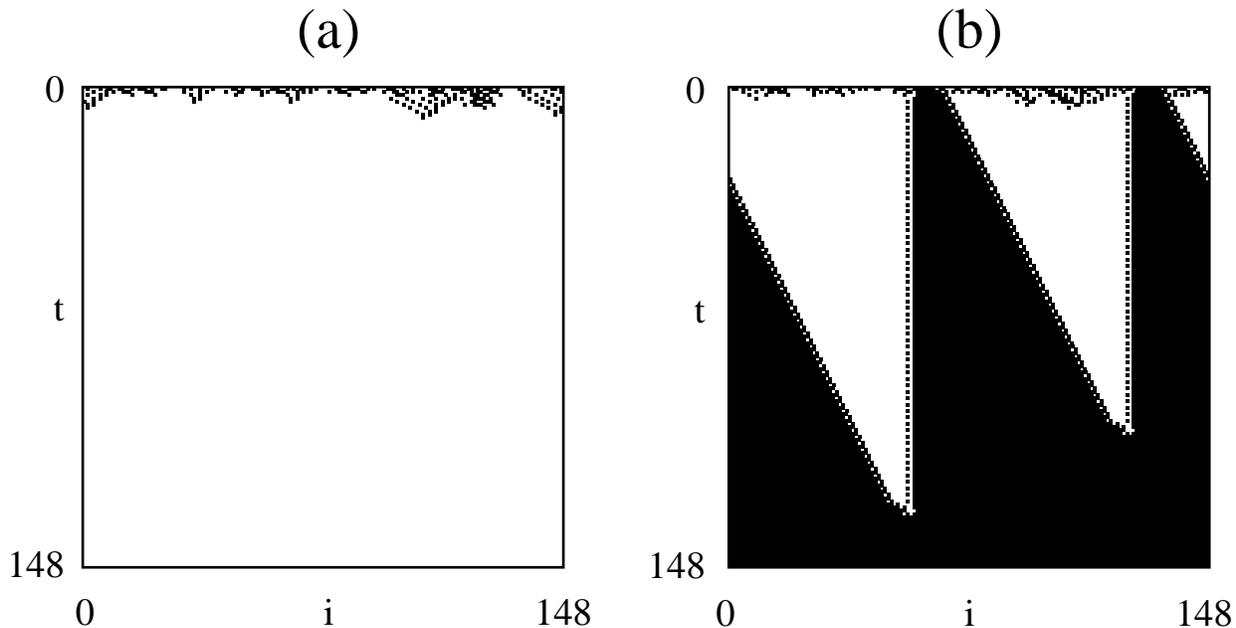

Figure 15: Space-time diagrams of one epoch-4 rule with $\lambda \approx 0.38$ that increases sufficiently large blocks of adjacent or nearly adjacent 1's. In (a) $\rho(0) \approx 0.44$; in (b) $\rho(0) \approx 0.52$. Both initial configurations are correctly classified.

The increase in fitness seen in Figure 11 between generation 10 and 20 or so is due to further refinements of the basic strategies that correct these problems to some extent.

**Epoch 4: Reaching and staying at a maximal fitness**

In most runs, the best fitness is typically at its maximum value of 0.90 to 0.95 by generation 40 or so. In Figure 11 this occurs at approximately generation 20, and is marked as the beginning of epoch 4. The best fitness does not increase significantly after this; the GA simply finds a number of variations of the best strategies that all have roughly the same fitness. When we extended 16 of the 30 runs to 300 generations, we did not see any significant increase in the best fitness.

The actions of the best rules from generation 100 of two separate runs are shown in Figures 15 and 16. The leftmost space-time diagrams in each figure are for initial configurations with $\rho < \rho_c$, and the rightmost diagrams are for initial configurations with $\rho > \rho_c$. The rule illustrated in Figure 15 has $\lambda = 0.38$; its strategy is to map initial configurations to 0's unless there is a sufficiently large block of adjacent or nearly adjacent 1's, which if present is increased. The rule shown in Figure 16 has $\lambda = 0.59$ and has the opposite strategy. Each of these rules has fitness $\approx 0.93$. They are better tuned versions of the rules in Figures 12 and 13.

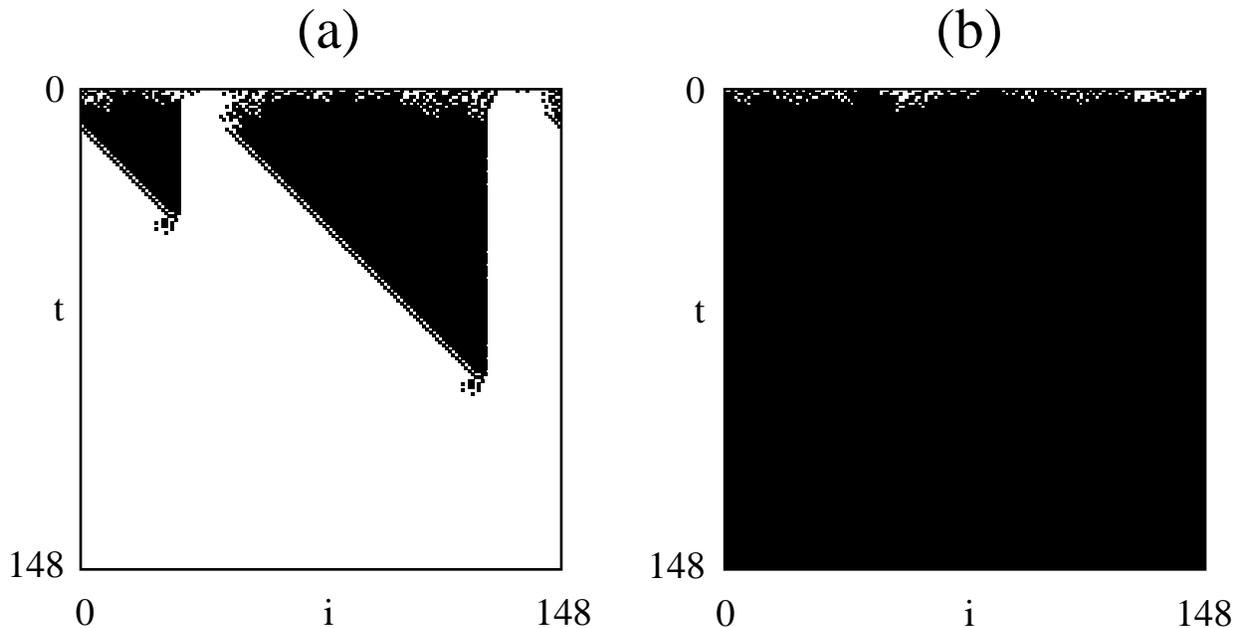

Figure 16: Space-time diagrams of one epoch-4 rule with $\lambda \approx 0.59$ that increases sufficiently large blocks of adjacent or nearly adjacent 0's. In (a) $\rho(0) \approx 0.40$; in (b) $\rho(0) \approx 0.56$. Both initial configurations are correctly classified.

**Symmetry breaking in epoch 3**

Notice that the $\lambda$ values of the rules that have been described are in the bins centered around 0.43 and 0.57 rather than 0.5. In fact, it seems to be much easier for the GA to discover versions of the successful strategies close to $\lambda = 0.43$ and $\lambda = 0.57$ than to discover them close to $\lambda = 1/2$, though some instances of the latter rules were found. Why is this? One reason is that rules with high or low $\lambda$ work well by *specializing*. The rules with low $\lambda$ map most neighborhoods to 0's and then increase sufficiently large blocks of 1's when they appear. Rules with high $\lambda$ specialize in the opposite direction. A rule at $\lambda = 1/2$ cannot easily specialize in this way. Another reason is that a successful rule that grows sufficiently large blocks of (say) 1's must avoid *creating* a sufficiently large block of 1's from an initial configuration with less than half 1's. Doing so will lead it to increase the block of 1's and produce an incorrect answer, as was seen in Figure 14. An easy way for a rule to avoid creating a sufficiently large block of 1's is to have a low $\lambda$. This ensures that low-density initial configurations will quickly map to all 0's, as was seen in Figure 15. Likewise, if a rule increases sufficiently large blocks of 0's, it is safer for the rule to have a high $\lambda$ value so it will avoid creating sufficiently large blocks of 0's where none existed. A rule close to $\lambda = 1/2$ will not have this safety margin, and may be more likely to inadvertently create a block of 0's or 1's that will lead it to a wrong answer. A final element that contributes to the difficulty of finding good rules with $\lambda = 1/2$ is the combinatorially large number of rules there. In effect, the search space is much larger, which makes the global search more difficult. Locally, about a given adequate rule at $\lambda = 1/2$, there are many more rules close in Hamming distance and thus reachable via mutation that are not markedly better.

Once the more successful versions of the epoch-3 strategies are discovered in epoch 4, their variants spread in the population, and the most successful rules have $\lambda$ on the low or high side of $\lambda = 1/2$. This explains the shift from the clustering around $\lambda = 1/2$ as seen in generations 10–30 in Figure 10 to a two-peaked distribution that becomes clear around generation 65. The rules in each run cluster around one or the other peak, specializing in one or the other way. We believe this type of symmetry breaking is a key mechanism that determines much of the population dynamics and the GA's success—or lack thereof—in optimization.

How does this analysis of the symmetry breaking jibe with the argument given earlier that the best rules for the $\rho_c = 1/2$ task must be close to $\lambda = 1/2$? None of the rules found by the GA had a fitness as high as 0.98—the fitness of the GKL rule, whose $\lambda$ is exactly $1/2$. That is, the evolved rules make significantly more classification errors than the GKL rule. To obtain the fitness of the GKL rule a number of careful balances in the rule table must be achieved. This is evidently very hard for the GA to do, especially in light of the symmetries in the task and their suboptimal breaking by the GA.

## 7.5 Performance of the Evolved Rules

Recall that the proportional fitness of a rule is the fraction of correct cell states at the final time step, averaged over 300 initial configurations. This fitness gives a rule partial credit for getting some final cell states correct. However, the actual task is to relax to either all 1's or all 0's, depending on the initial configuration. In order to measure how well the evolved rules actually perform the task, we define the *performance* of a rule to be the fraction of times the rule correctly classifies initial configurations, averaged over a large number of initial configurations. Here, credit is given only if the initial configuration relaxes to exactly the correct fixed point after some number of time steps. We measured the performance of each of the elite rules in the final generations of the 30 runs by testing it on 300 randomly generated initial configurations that were uniformly distributed in the interval $0 \leq \rho \leq 1$, letting the rule iterate on each initial condition for 1000 time steps. Figure 17 displays the mean performance (diamonds) and best performance (squares) in each $\lambda$ bin. This figure shows that while the mean performances in each bin are much lower than the mean fitnesses for the elite rules shown in Figure 7, the best performance in each bin is roughly the same as the best fitness in that bin. (In some cases the best performance in a bin is slightly higher than the best fitness shown in Figure 7. This is because different sets of 300 initial conditions were used to calculate fitness and performance. This difference can produce small variations in the fitness or performance values.) The best performance we measured was $\approx 0.95$. Under this measure the performance of the GKL rule is $\approx 0.98$. Thus the GA never discovered a rule that performed as well as the GKL rule, even up to 300 generations. In addition, when we measure the performance of the fittest evolved rules on larger lattice sizes, their performances decrease significantly, while that of the GKL rule remains roughly the same.

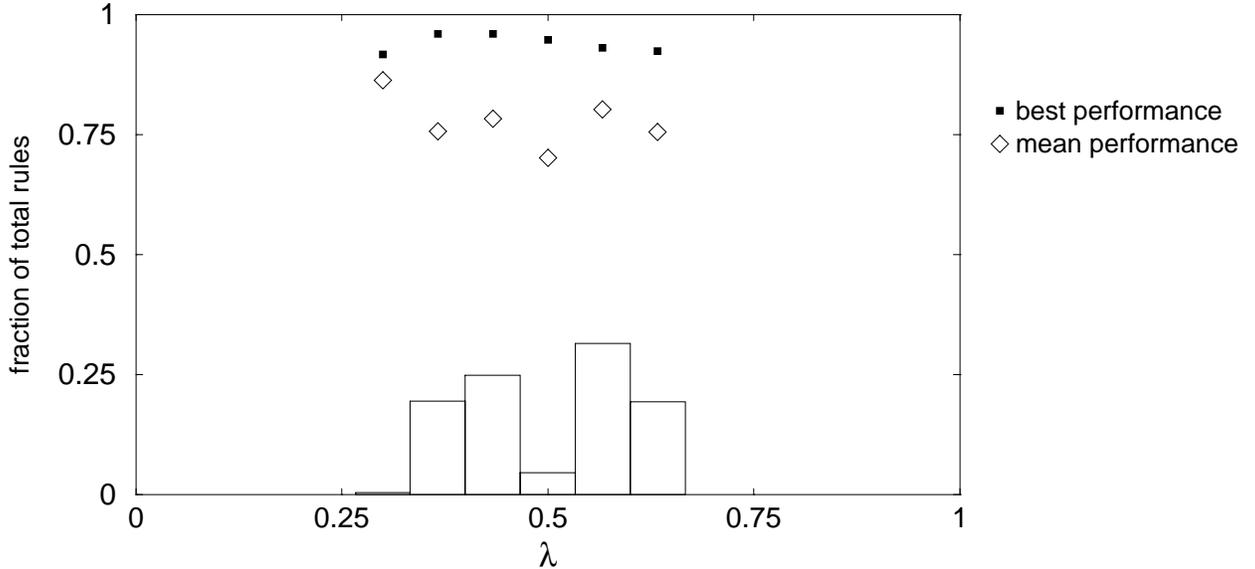

Figure 17: Performances of the rules evolved with the proportional-fitness function. Only the elite rules from generation 100 are included in this histogram. The mean performance in each bin (open diamonds), and the best performance in each bin (black squares) is plotted.

## 7.6 Using Performance as the Fitness Criterion

Can the GA evolve better-performing rules on this task? To test this, we carried out an additional experiment in which performance as defined in the previous section is the fitness criterion. As before, at each generation each rule is tested on 300 initial configurations that are uniformly distributed over density values. However, in this experiment, a rule's fitness is the fraction of initial configurations that are correctly classified. An initial configuration is considered to be incorrectly classified if any bits in the final lattice are incorrect. Aside from this modified fitness function, everything about the GA remained the same as in the proportional-fitness experiments. We performed 30 runs of the GA for 100 generations each. The results are given in Figure 18, which gives a histogram plotting the frequencies of the elite rules from generation 100 of all 30 runs as a function of $\lambda$. As can be seen, the shape of the histogram again has two peaks centered around a dip at $\lambda = 1/2$. This shape results from the same symmetry-breaking effect that occurred in the proportional-fitness case: these runs also evolved essentially the same strategies as the epoch-3 strategies described earlier. The best and mean performances here are comparable to the best performances in the proportional-fitness case; the best performances found here are $\approx 0.95$. The performance as a function of $\rho(0)$ for one of the best rules is plotted in Figure 19, for lattice sizes of 149 (the lattice size used for testing the rules in the GA runs), 599, and 999. We used the same procedure to make these plots as was described earlier for Figure 4. As can be seen, the performance according to this measure is significantly worse than that of the GKL rule (cf. Figure 4), especially on larger lattice sizes. The worst performances for all lattice sizes are centered close to the rule's $\lambda$ value of $\approx 0.42$.

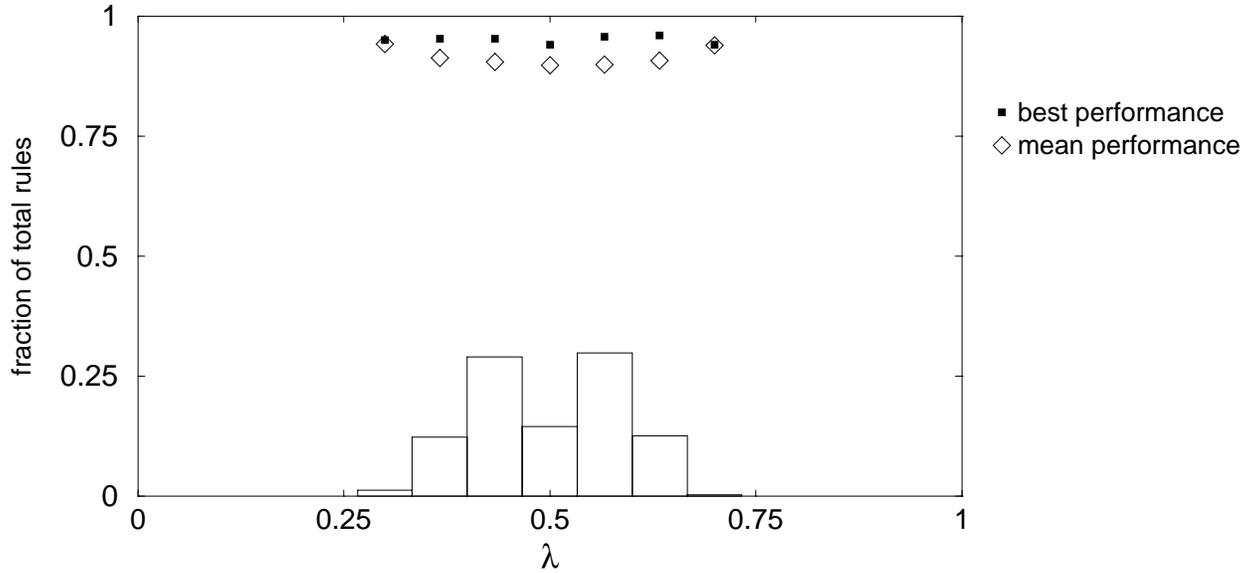

Figure 18: Results from our experiment with performance as the fitness criterion. The histogram plots the frequencies of elite rules merged from the final generations (generation 100) of 30 runs in which the performance-fitness function was used.

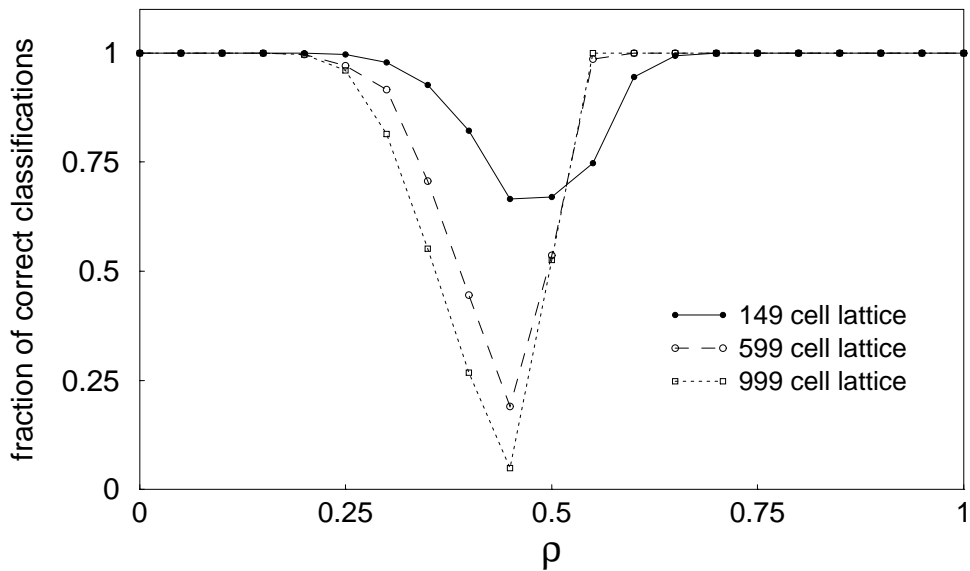

Figure 19: Performance of one of the best rules evolved using performance fitness, plotted as a function of $\rho(0)$. Performance plots are given for three lattice sizes: 149 (the size of the lattice used in the GA runs), 599, and 999. This rule has $\lambda \approx 0.42$.

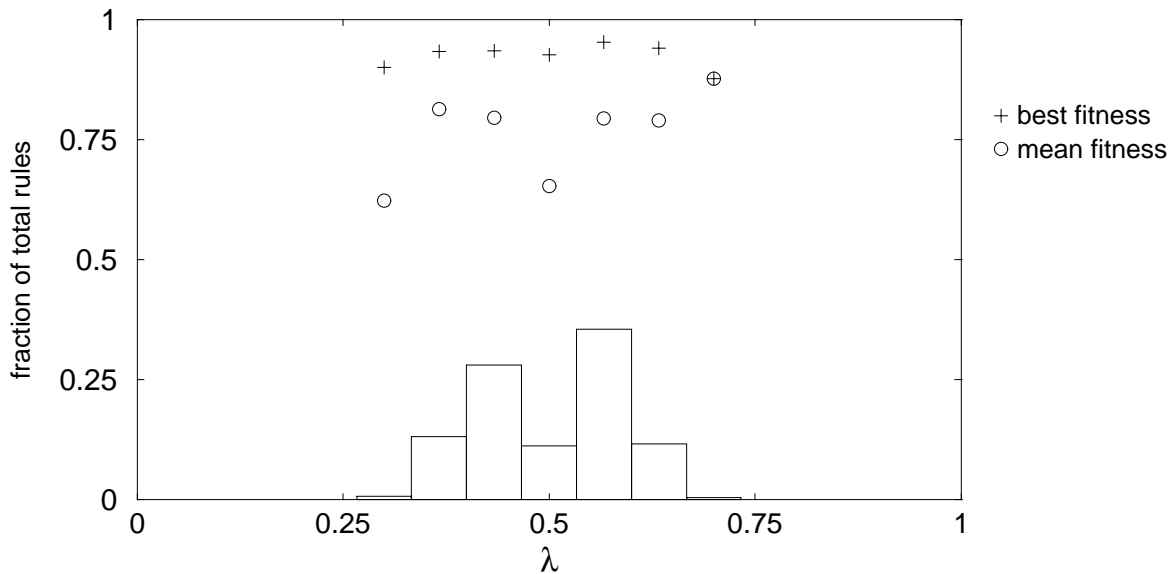

Figure 20: Results from our experiment in which a diversity-enforcement mechanism was added to the GA. The histogram plots the frequencies of rules merged from the entire population at generation 100 of 20 runs in which our diversity-enforcement scheme was used.

## 7.7 Adding A Diversity-Enforcement Mechanism

The description given above of the four epochs in the GA's search explains the results of our experiment, but it does not explain the difference between our results and those of the original experiment reported in [23]. One difference between our GA and the original was the inclusion in the original of a diversity-enforcement scheme that penalized newly formed rules that were too similar in Hamming distance to existing rules in the population. To test the effect of this scheme on our results, in one set of experiments we included a similar scheme. In our scheme, every time a new string is created through crossover and mutation, the average Hamming distance between the new string and the elite strings—the 50 strings that are copied unchanged—is measured. If this average distance is less than 30% of the string length (here 38 bits), then the new string is not allowed in the new population. New strings continue to be created through crossover and mutation until 50 new strings have met this diversity criterion. We note that many other diversity-enforcement schemes have been developed in the GA literature; e.g., "crowding" [8].

The results of this experiment are given in Figure 20. The histogram in that figure represents the merged rules from the entire population at generation 100 of 20 runs of the GA, using the proportional-fitness function and our diversity-enforcement scheme. As can be seen, the histogram in this figure is very similar to that in Figure 6(b). The only major difference is the significantly lower mean fitness in the middle and leftmost bins, which results from the increased requirement for diversity in the final non-elite population. We conclude that the use of this diversity-enforcement scheme was not responsible for the difference between the results from [23] and our results.

## 7.8 Differences Between Our Results and the Original Experiment

As was seen in Figure 6(b), our results are strikingly different from those reported in [23]. These experimental results, along with the theoretical argument that the most successful rules for this task should have $\lambda$ close to $1/2$, lead us to conclude that the interpretation of the original results as giving evidence for the hypotheses concerning evolution, computation, and $\lambda$ is not correct. However, we do not know what accounted for the differences between our results and those obtained in the original experiment. We speculate that the differences are due to additional mechanisms in the GA used in the original experiment that were not reported in [23].

Although the results were very different, there is one qualitative similarity: the rule-frequency-versus-$\lambda$ histograms in both cases contained two peaks separated by a dip in the center. As already noted, in our histogram the two peaks were closer to $\lambda = 1/2$ by a factor of 4, but it is possible that the original results were due to a mechanism similar to (i) the epoch-0 sensitivity to initial configuration and population asymmetry about $\lambda = 1/2$ or (ii) the symmetry breaking we observed in epoch 3, as described above. Perhaps these were combined with some additional force in the original GA that kept rules far away from $\lambda = 1/2$. Unfortunately, the best and mean fitnesses for the $\lambda$ bins were not reported for the original experiment. As a consequence we do not know whether or not the peaks in the original histogram contained high-fitness rules, or even if they contained rules that were more fit than rules in other bins. Our results and the basic symmetry in the problem suggest otherwise.

## 8. General Discussion

### 8.1 What We Have Shown

The results reported in this paper have demonstrated that the results from Packard's original experiment do not hold up under our experiments. We conclude that the original experiment does not give firm evidence for the hypotheses it was meant to test: first, that rules capable of performing complex computation are most likely to be found close to $\lambda_c$ values and, second, that when CA rules are evolved by a GA to perform a nontrivial computation, evolution will tend to select rules close to $\lambda_c$ values.

As we argued theoretically and as our experimental results suggest, the most successful rules for performing a given $\rho$-classification task will be close to a particular value of $\lambda$ that depends on the particular $\rho_c$ of the task. Thus for this class of computational tasks, the $\lambda_c$ values associated with an "edge of chaos" are not correlated with the ability of rules to perform the task.

In the remainder of this section, we step back from these particular experiments and discuss in more general terms the ideas that motivated these studies.

## 8.2 $\lambda$, Dynamical Behavior, and Computation

As was noted earlier, Langton presented evidence that, given certain caveats regarding the radius $r$ and number of states $K$, there is some correlation between $\lambda$ and the behavior of an "average" CA on an "average" initial configuration [16]. Behavior was characterized in terms of quantities such as single-site entropy, two-site mutual information, difference-pattern spreading rate, and average transient length. The correlation is quite good for very low and very high $\lambda$ values, which predict fixed-point or short-period behavior. However, for intermediate $\lambda$ values, there is a large degree of variation in behavior. Moreover, there is no precise correlation between these $\lambda$ values and the location of a behavioral "phase transition", other than that described by Wootters and Langton in the limit of infinite $K$.

The remarks above and all the experimental results in [16] are concerned with the relationship between $\lambda$ and the dynamical behavior of CA. They do not directly address the relationship between $\lambda$ and computational capability of CA. The basic hypothesis was that $\lambda$ correlates with computational capability in that rules capable of complex, and in particular, universal, computation must be, or at least are most likely to be, found near $\lambda_c$ values. As far as CA are concerned,[10] the hypothesis was based on the intuition that complex computation cannot be supported in the short-period or chaotic regimes because the phenomena that apparently occur only in the "complex" (non-periodic, non-chaotic) regimes, such as long transients and long space-time correlation, are necessary to support complex computation. There has thus far been no experimental evidence correlating $\lambda$ with an independent measure of computation. Packard's experiments were intended to address this issue since they involved an independent measure of computation—performance on a particular complex computational task—but as we have shown, these experiments do not provide evidence for the hypothesis linking $\lambda_c$ values with computational ability. In order to test this hypothesis, more general measures of computation need to be used.

The argument that complex computation cannot occur in the short-period or chaotic regimes may seem intuitively correct, but there is actually a theoretical framework and strong experimental evidence to the contrary. Hanson and Crutchfield [2, 11] have developed a method for filtering out chaotic "domains" in the space-time diagram of a CA, sometimes revealing "particles" that have the non-periodic, non-chaotic properties of structures in Wolfram's Class 4 CA. That is, with the appropriate filter applied, complex structures can be uncovered in a space-time diagram that, to the human eye and to the statistics used in [16], appears to be completely random. As an extreme example, it is conceivable that such filters could be applied to a seemingly chaotic CA and reveal that the CA is actually implementing a universal computer (with glider guns implementing AND, OR, and NOT gates, etc.). Hanson and Crutchfield's results strikingly illustrate that apparent complexity of behavior—and apparent computational capability—can depend on the implicit "filter" imposed by one's chosen statistics.

---

[10] In the context of continuous-state dynamical systems, it has been shown that there is a direct relationship between intrinsic computational capability of a process and the degree of randomness of that process at the phase transition from order to chaos. Computational capability was quantified with the statistical complexity, a measure of the amount of memory of a process, and via the detection of an embedded computational mechanism equivalent to a stack automaton.[3]

## 8.3 What Kind of Computation in CA Do We Care About?

In the section above, the phrases "complex computation" and "computational capability" were used somewhat loosely. As was discussed in Section 3, there are at least three different interpretations of the notion of computation in CA. The notion of a CA being able to perform a "complex computation" such as the $\rho_c = 1/2$ task, where the CA performs the same computation on all initial configurations, is very different from the notion of a CA being capable, under some special set of initial configurations, of simulating a universal computer. Langton's speculations regarding the relationship between dynamical behavior and computational capability seemed to be more concerned with the latter than the former, though the implication is that the capability to sustain long transients, long correlation lengths, and so on are necessary for both notions of computation.

If "computationally capable" is taken to mean "capable, under some initial configuration(s), of universal computation", then one might ask why this is a particularly important property of CA on which to focus. In [16] CA were used merely as a vehicle to study the relationship between phase transitions and computation, with an emphasis on universal computation. But for those who want to use CA as scientific models or as practical computational tools, a focus on the capacity for universal computation may be misguided. If a CA is being used as a model of a natural process (e.g., turbulence), then it is currently of limited interest to know whether or not the process is in principle capable of universal computation if universal computation will arise only under some specially engineered initial configuration that the natural process is extremely unlikely to ever encounter. Instead, if one wants to understand emergent computation in natural phenomena as modeled by CA, then one should try to understand what computation the CA "intrinsically" does [2, 11] rather than what it is "in principle capable" of doing only under some very special initial configurations. Thus, understanding the conditions under which a capacity for universal computation is possible will not be of much value in understanding the natural systems modeled by CA.

This general point is neither new nor deep. Analogous arguments have been put forth in the context of neural networks, for example. While many constructions have been made of universal computation in neural networks (e.g., [28]), some psychologists (e.g., [27]) have argued that this has little to do with understanding how brains or minds work in the natural world.

Similarly, if one wants to use a CA as a parallel computer for solving a real problem—such as face recognition—it would be very inefficient, if not practically impossible, to solve the problem by (say) programming Conway's Game of Life CA to be a universal computer that simulates the action of the desired face recognizer. Thus understanding the conditions under which universal computation is possible in CA is not of much practical value either.

In addition, it is not clear that anything like a drive toward universal-computational capabilities is an important force in the evolution of biological organisms. It seems likely that substantially less computationally-capable properties play a more frequent and robust role. Thus asking under what the conditions evolution will create entities (including CA) capable of universal computation may not be of great importance in understanding natural evolutionary mechanisms.

In short, it is mathematically important to know that some CA are in principle capable of universal computation. But we argue that this is by no means the most scientifically interesting property of CA. More to the point, this property does not help scientists much in understanding the emergence of complexity in nature or in harnessing the computational capabilities of CA to solve real problems.

## 9. Conclusion

The main purpose of this study was to examine and clarify the evidence for various hypotheses related to evolution, dynamics, computation, and cellular automata. We hope this study has shed some new and constructive light on these issues. As a result of our study we have identified a number of evolutionary mechanisms, such as the role of combinatorial drift, and the role of symmetry and the impediments to emerging computational strategies caused by symmetry breaking. For example, we have found that the breaking of the goal task's symmetries in the early generations can be an impediment to further optimization of individuals in the population. The symmetry breaking results is a kind of suboptimal speciation in the population that is stable or, at least, meta-stable over long times. The symmetry-breaking effects we described here may be similar to symmetry-breaking phenomena such as bilateral symmetry and handedness that emerge in biological evolution. It is our goal to develop a more rigorous framework for understanding these mechanisms in the context of evolving CA. We believe that a deep understanding of these mechanisms in this relatively simple context can yield insights for understanding evolutionary processes in general and for successfully applying evolutionary-computation methods to complex problems.

Though our experiments did not reproduce the results reported in [23], we believe that the original conception of using GAs to evolve computation in CA is an important idea. Aside from its potential for studying various theoretical issues, it also has a potential practical side that could be significant. As was mentioned earlier, CA are increasingly being studied as a class of efficient parallel computers; the main bottleneck in applying CA more widely to parallel computation is *programming*—in general it is very difficult to program CA to perform complex tasks. Our results suggest that the GA has promise as a method for accomplishing such programming automatically. In order to further test the GA's effectiveness as compared with other search methods, we performed an additional experiment, comparing the performance of our GA on the $\rho_c = 1/2$ task with the performance of a simple steepest-ascent hill-climbing method. We found that the GA significantly outperformed hill-climbing, reaching much higher fitnesses for an equivalent number of fitness evaluations. This gives some evidence for the relative effectiveness of GAs as compared with simple gradient ascent methods for programming CA. Koza [15] has also evolved CA rules using a very different type of representation scheme; it is a topic of substantial practical interest to study the relationship of representation and GA success on such tasks.

### Acknowlegments

This research was supported by the Santa Fe Institute, under the Adaptive Computation and External Faculty Programs, and by the University of California, Berkeley, under contract AFOSR 91-0293. Thanks to Doyne Farmer, Jim Hanson, Erica Jen, Chris Langton, Wentian Li, Cris Moore, and Norman Packard for many helpful discussions and suggestions concerning this project. Thanks also to Emily Dickinson and Terry Jones for technical advice.